\documentstyle[epsf]{mn}
\title[Cool White Dwarf Survey]{A Survey for Cool White Dwarfs and
the age of the Galactic Disc}
\author[R.A.\ Knox, M.R.S.\ Hawkins and N.C.\ Hambly]{R.A.\ Knox$^1$,
 M.R.S.\ Hawkins$^{1,2}$, N.C.\ Hambly$^{1,2}$ \\
$^1$Institute for Astronomy, University of Edinburgh,
 Blackford Hill, Edinburgh,EH9~3HJ
\\
$^2$Royal Observatory, Blackford Hill, Edinburgh, EH9~3HJ}

\date{Accepted ---. Received ---; in original form ---}

\begin{document}
\maketitle
\begin{abstract}
We describe a new multi--colour proper motion survey for cool white dwarfs
(CWDs).  The observational database consists of $\sim300$ digitally scanned
Schmidt plates in ESO/SERC field 287.
The entire survey procedure, from the raw Schmidt plate data to final white dwarf
luminosity function (WDLF) is described, with special emphasis on completeness 
concerns.

We obtain a sample of 58 WDs, for which we have follow up CCD photometry and 
spectroscopy of a representative sub--sample.  Effective temperatures and 
luminosities of our sample objects are determined by comparing photometry with
the model atmosphere predictions of Bergeron, Saumon and Wesemael.  Space densities
are calculated using an adaptation of Schmidts $(1/V_{max})$ method, from which
a WDLF is constructed.  Comparison of our observational LF with the models of both Wood
and Garc\'{\i}a-Berro et al. indicate an age
 for the local Galactic Disc of $10^{-1}_{+3}$~Gyr.
Importantly, we find no evidence of incompleteness in our survey sample.  Proper motion
number counts imply the survey is complete, and the WD sample passes
the $(V/V_{max})$ completeness test.
\end{abstract}

\section{Introduction}

A reliable age for the local Galactic Disc places a valuable constraint on
both globular cluster ages and cosmological models.  A number of 
independent methods
of investigating this problem have been employed in the past (eg. Jimenez 1998
and references therein), resulting in a broad consensus that the lower
limit for the Disc age lies between 8 and 12~Gyr.
Potentially one of the most reliable means of estimating the Disc age is via
cool white dwarf (CWD) stars.
These estimates use the idea, first proposed by Schmidt (1959), that in a galaxy
of finite age there will be a temperature beyond which the oldest, coolest
white dwarfs
(WDs) have not had time to cool.  This predicted cut-off in the luminosity function (LF)
of WDs, if satisfactorily observed, can then be used in conjunction with
WD cooling models to derive the Disc age.

CWDs are difficult to find, being both extremely faint and of similar
colour to the numerous K and M-type dwarfs; and are almost exclusively discovered
by means of their proper motion.  The cut-off in the WDLF was observed 
(Liebert, Dahn \& Monet 1988, hereafter LDM) after
thorough follow up observations of CWD candidates drawn from the Luyten
Half Second (LHS) Catalogue (Luyten 1979).  Although at that time a Disc age
of $9.3\pm2$~Gyr was derived from this sample (Winget et al. 1987), 
further observations and improvements in model atmospheres 
(Bergeron, Ruiz \& Leggett 1997, hereafter BRL) and theoretical LFs 
(Wood 1992, 1995) has prompted a recent 
redetermination of the Disc age for the same sample (Leggett,
Ruiz \& Bergeron 1998, hereafter LRB), yielding a value of $8\pm1.5$~Gyr.
While the existence of the cut-off in the LDM WDLF has not been challenged by
subsequent observational work, the details of its precise position and shape have.
A sample of CWDs found using common proper motion binaries (CPMBs), again culled
from the Luyten surveys, suggest that there are $\sim5$ times more very
faint CWDs than found by LDM (Oswalt et al. 1996, hereafter OSWH).
A Disc age of $9.5\pm1$~Gyr was found using this sample, and the factor 
of $\sim5$ increase in the faintest WDs has been confirmed by an independent
search for CWDs in the south (Ruiz and Takamiya 1995).

Until now, the proper motion catalogues used to extract samples of CWDs have
been produced by `blink' comparison.  While these surveys have clearly been 
successful in picking up individual stars of low luminosity and high proper motion,
the use of such a subjective survey technique raises worries concerning completeness.
The advent of high precision micro-densitometers such as SuperCOSMOS (Hambly 1998 
and references therein) allow proper motions and magnitudes to be calculated
objectively using a series of plates in the same field.  Hambly et al. (1997) have
recently discovered in Taurus perhaps the coolest known WD using
just such a procedure.  This object, WD 0346+246, should certainly appear
 in the LHS Catalogue, and would then presumably have been included in the
 LDM sample since it fulfills the given criteria. This object was discovered
serendipitously from work in a particular $6\times6^\circ$ Schmidt field, and therefore
casts further doubt on the completeness of the Luyten catalogue.

In this work we exploit both the 
extensive collection of over 300 plates in ESO/SERC field 287 and the power of the 
SuperCOSMOS measuring machine to produce a deep, complete multi-colour proper
 motion survey from which a sample of CWDs has been extracted.  
Much attention was given to the critical question of completeness, with specific
reference to the choice of survey limits and the sources of potential contaminants.
The paper is organised as follows: the plate database and reduction of the 
digitally scanned data is described in Section~\ref{schmidt}; the choice of survey
limits to combat sample contamination and the method of WD sample selection
are discussed in Section~\ref{limits}; in Section~\ref{hipms} the important
question of the high proper motion limit is addressed by independent tests; some
follow up observations of our WD candidates are presented in  Section~\ref{followup};
stellar parameters are calculated for our sample in Section~\ref{samanal}, which also
includes a discussion of potential contaminant populations; the WDLF from our sample
is presented in Section~\ref{wdlf}, and the derived Disc ages discussed in 
Section~\ref{discuss}.  

\section{Schmidt Plate Data Reduction}
\label{schmidt}

\begin{figure}
\begin{center}
\setlength{\unitlength}{1mm}
\begin{picture}(70,40)
\includegraphics{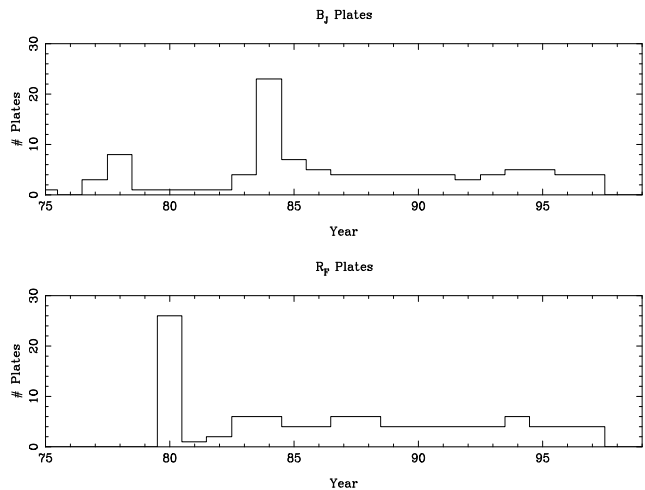} 
\end{picture}
\end{center}
\vspace{2cm}
\caption{Number of plates per year in ESO/SERC Field 287 in the $\rm B_{J}$ 
(top) and $\rm R_{F}$(bottom) passbands}
\label{platehist}
\end{figure}

In the course of a long term quasar variability study (eg. V\'{e}ron \& Hawkins 1995) over
300 Schmidt plates have been taken in ESO/SERC field 287.  Figure~\ref{platehist}
shows the distribution over time of the plate collection in this field
in the two principal passbands available, $\rm B_{J}$ and $\rm R_{F}$
.  Additional plate material exists
in the U (19 plates), V (11 plates) and I (40 plates) passbands.  All these
plates have been digitally scanned by either the SuperCOSMOS plate measuring 
machine or its predecessor COSMOS
 (MacGillivray and Stobie, 1984).
As described in Appendix~\ref{rpms}, a
 CWD survey utilising the reduced proper motion (RPM)
 population discrimination technique requires
photometry in at least two passbands and proper motion data.  Although
for most stellar types a survey of this sort will be proper motion limited,
for extremely intrinsically faint objects such as cool degenerates the
photometric survey limits also become important. A principal concern in this
work is therefore to maximise the survey depth.

  The technique of stacking
digitised Schmidt plates has been in existence for some time
(eg.\ Hawkins~1991; Kemp \& Meaburn~1993; Schwartzenberg et al.~1996),
and has recently been thoroughly investigated with SuperCOSMOS 
(Knox et al. 1998).  The database presented in Figure~\ref{platehist} is an
obvious candidate for stacking in a proper motion survey, since at least 4 plates
are available in most years and the use of 4 plate stacks rather than
single plate data will yield deeper photometric survey limits with no
reduction in the proper motion time baseline.  The data has therefore been
grouped into 4 plate stacks according to Table~\ref{epochs}.
\begin{table}
\begin{center}
\begin{tabular}{|ccc|ccc|ccc|} \hline\hline
Year&$\rm B_{J}$&R  &Year&$\rm B_{J}$&R  &Year&$\rm B_{J}$&R  \\ \hline
1977& 1 &   &1984& 4 & 2 &1991& 1 & 1 \\
1978& 1 &   &1985& 2 & 1 &1992& 1 & 1 \\
1979&   &   &1986& 1 & 1 &1993& 1 & 1 \\
1980&   & 5 &1987& 1 & 1 &1994& 1 & 1 \\
1981&   &   &1988& 1 & 1 &1995& 1 & 1 \\
1982&   &   &1989& 1 & 1 &1996& 1 & 1 \\
1983& 1 & 1 &1990& 1 & 1 & & & \\
\hline\hline
\end{tabular}
 \caption{The number of $\rm B_{J}$ and R 4 plate stacks per annual epoch.}
 \label{epochs}
\end{center}
\end{table}

In order to track objects through epochs and calculate relative 
proper motions, all the
stacks have been shifted to a common co-ordinate system corresponding to 
a measure of the high quality $\rm B_{J}$ plate J3376.
This is achieved by a global 
transformation followed by local transformations (translation, rotation and
scale), performed by splitting the 
stack area into a grid of 16x16 small areas (Hawkins 1986).
  Every object in each area is used
to define the local transformation (since fainter objects are more numerous,
they essentially define the astrometric coordinate system).  Objects are then
paired up between epochs.  The software used in this work operates using
a fixed `box size' (13arcecs) in which it looks for a pair, thus in the
first instance a high proper motion limit is imposed on the survey -- 
this issue will be addressed in detail later.
The pairing procedure yields a list of x and y coordinates for 
each object, one set for each 
stack the object is found on. Calculating proper motions is 
then simply a matter of performing a linear regression fit to each object's
x and y coordinates as a function of time.  However, erroneous pairing 
inevitably occurs between stacks, and we therefore wish to perform some form 
of bad point rejection to reduce contamination by spurious proper motions.
In order to reject deviant points (and calculate parameters such as $\sigma
_{\mu}$ and $\chi^{2}$) an estimate of the error associated with each measure
of position is required.  We assume this error is simply a function of
 magnitude
and that it will vary from stack to stack, but not across the survey area.
This error is calculated using the deviation of an object's position
on a particular stack from the mean position over the 20 stacks used, and
is determined over 10 magnitude bins.
A $3\sigma$ iterative rejection procedure is implemented to reject spurious
pairings or high proper motion objects which are not reflecting the true
positional errors sought.   The calculated 
errors are much as one might expect: decreasing for
brighter objects until factors such as saturation
 and blended images makes positional measures more uncertain.

\begin{figure}
\begin{center}
\setlength{\unitlength}{1mm}
\begin{picture}(70,40)
\includegraphics{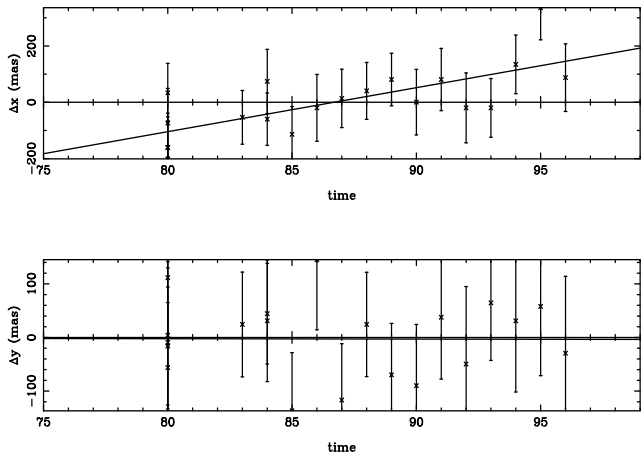} 
\end{picture}
\end{center}
\vspace{2cm}
\caption{An example of a linear regression fit to a set of object positions.
The vertical axis is deviation of measured position on a given stack from the
mean object position, with year this century on the horizontal axis} 
 \label{pmplot}
\end{figure}

A straight line fit can now be applied to the x and y data for each object,
the gradient of which is taken to be the measured proper motion, $\mu_{x}$ and 
$\mu_{y}$ respectively.  An example
is shown in Figure~\ref{pmplot}, the points showing the deviation at each
 epoch from
the average object position with error bars calculated as above.
Deviant points arising from spurious pairings often
 lie far from the other data and will give 
rise to spurious high proper motion detections if not removed.  We therefore
iteratively remove points lying $3\sigma$ from the fitted line.
This can occasionally lead to further problems if there are several bad points
associated with the object, and the result of several iterations can be a
 larger spurious motion detection.  This source of contamination is generally
eliminated by insisting sample objects are detected on virtually every stack.
The validity of the  positional error estimation scheme described
above has been verified by confirming that scatter plots of log reduced
$\chi^{2}$ as a function of magnitude cluster around zero for all magnitudes 
in all regions of the survey area.

Instrumental magnitudes are calculated for every object detected
on each stack in the standard COSMOS/SuperCOSMOS fashion (Beard et al. 1990 and
references therein).  Briefly, an object detection is defined by a given number
of interconnected pixels with intensity above a given threshold 
(eg. 8 interconnected pixels with intensity above a $2.5\sigma$  sky noise
threshold for SuperCOSMOS data).
An object's instrumental magnitude is then calculated as the log of the sum of 
the intensity above background across the object area.
This quantity varies monotonically with true magnitude, and is therefore
suitable for use in constructing calibration curves using a CCD sequence.
A sequence of $\sim 200$ stars with CCD magnitudes measured in a variety of
passbands exists in field  287 (Hawkins et al. 1998)
, yielding U, B, V, R and I photometry to a typical
accuracy of 0.15 magnitudes (see Section~\ref{photom}).  Significantly smaller
errors are theoretically obtainable from photographic material, and the larger
uncertainties we find appear to be caused by systematic deviations of sequence objects
from the calibration curve.  This is not a colour or field effect, and is probably caused
by differences in detection media.
 
\section{Survey limits and Sample Selection}
\label{limits}

The `catalogue' resulting from the implementation of the procedure described
in the previous section consists of astrometric and photometric measures
for over 200,000 objects.  Criteria for inclusion in this preliminary sample
is merely detection in both $\rm B_{J}$ and R passbands (since these are required
for construction of the reduced proper motion diagram (RPMD))
 and a measure of proper motion in both 
these passbands.  It is from these objects that an uncontaminated proper
motion sample is to be drawn; and we require well defined universal survey 
limits
so that space densities can be calculated from the final survey sample.

\begin{figure}
\begin{center}
\setlength{\unitlength}{1mm}
\begin{picture}(70,40)
\includegraphics{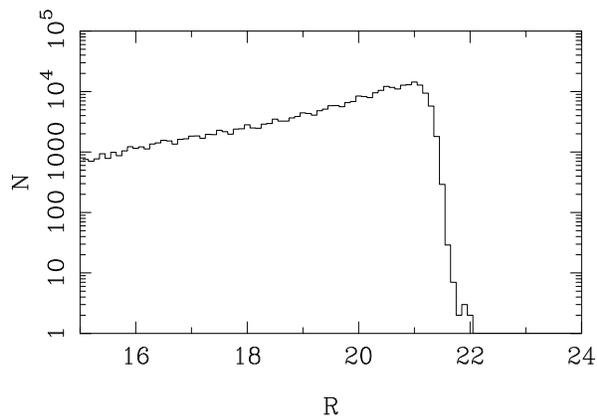} 
\end{picture}
\end{center}
\vspace{2cm}
\caption{Number counts as a function of apparent R magnitude}
 \label{Rhist}
\end{figure}

Number count plots from this survey data
increase linearly with increasing magnitude, as shown for 
the R data in Figure~\ref{Rhist},
before dropping precipitously.  This cut-off is attributed to 
the survey detection limit, and the position of the turnover is used
to determine photometric survey limits. The limits used are 21.2 in R and
22.5 in B.

\begin{figure}
\begin{center}
\setlength{\unitlength}{1mm}
\begin{picture}(70,40)
\includegraphics{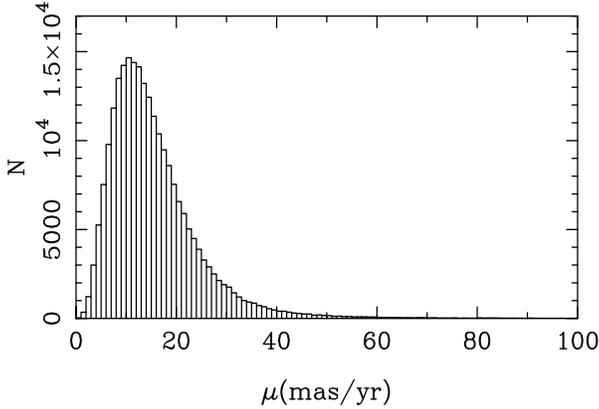} 
\end{picture}
\end{center}
\vspace{2cm}
\caption{Proper motion frequency distribution}
 \label{pmhist}
\end{figure}

The proper motion distribution for all objects in our survey area detected
on at least 15 stacks in both B and R is shown
in Figure~\ref{pmhist}.
Low proper motions are generally an artifact of measuring
machine error, thus the distribution indicates a typical error in measured proper motions of
$ \rm \sim10mas/yr$.  Our criteria for choosing a survey proper motion limit
are elimination of contaminant spurious proper motions from the sample and
, with this in mind, maximising the size of the final proper motion sample
extracted.
The deterioration of positional accuracy with magnitude means that 
fainter objects also have more uncertain proper motion determinations. The 
peak of the proper motion distribution therefore 
moves to higher proper motions with
object samples drawn from progressively fainter magnitude bins. For this
reason, we consider the survey proper motion limit as a function of
magnitude.

Three independent methods of determining an appropriate proper motion limit
have been investigated for this dataset:

\begin{enumerate}
 \renewcommand{\theenumi}{(\arabic{enumi})}
  \item  analysis of proper motion error distribution
  \item  cumulative proper motion number counts
  \item  RPMD inspection
\end{enumerate}

all of which are described here.

The characteristics of the `noise' in the proper motion
distribution can be analysed by assuming all objects are `zero proper motion 
objects' and the calculated proper motions arise purely from random 
measurement error.  The `true' values of \hbox{$\mu_{x}$} and \hbox{$\mu_{y}$} measured by
linear regression are therefore zero, and the random measurement errors give
rise to a normal error distribution in \hbox{$\mu_{x}$} and \hbox{$\mu_{y}$} about zero with common
$\sigma$.
We are 
of course 
interested in the total proper motion,
 which follows a Rayleigh distribution of the form
\begin{equation}
P(\mu)=\frac{\mu}{\sigma^{2}}e^{-\frac{\mu^{2}}{2\sigma^{2}}}.
\end{equation} 
This survey utilises two independent measures of proper motion in the same
field, one from the stacks of B plates and the other from the R stacks.  A
useful means of reducing the final proper motion survey limit will be to
compare these two measures and reject inconsistent motions.  Comparison of 
independent measures of proper motion needs to be incorporated into this
analysis if it is to be useful in predicting sample contamination later. 
The measured B and R motions and their associated error distributions are 
therefore characterised
\begin{equation}
\label{jdist}
P(\hbox{$\mu_{b}$}) = A_{b} \hbox{$\mu_{b}$} e^{- C_{b} \mu_{b}^{2}}
\end{equation}
and
\begin{equation}
\label{rdist}
P(\hbox{$\mu_{r}$}) = A_{r} \hbox{$\mu_{r}$} e^{- C_{r} \mu_{r}^{2}}
\end{equation}
respectively.
  The algorithm used to select sample objects on the basis of
proper motions will use three sequential cuts to eliminate spurious motions.
Firstly, the averaged proper motion must exceed the proper motion limit \hbox{$\mu_{lim}$}
, ie:
\begin{equation}
\frac{\hbox{$\mu_{r}$}+\hbox{$\mu_{b}$}}{2}>\hbox{$\mu_{lim}$}.
\end{equation}
Secondly, the difference in proper motion must not exceed a second survey 
parameter \hbox{$(\Delta\mu)$}:
\begin{equation}
|\hbox{$\mu_{b}$} - \hbox{$\mu_{r}$}| < \hbox{$(\Delta\mu)$}, 
\end{equation}
and finally the difference in position angle must not exceed a third survey 
parameter \hbox{$(\Delta\phi)$}:
\begin{equation}
|\hbox{$\phi_{j}$} - \hbox{$\phi_{r}$}| < \hbox{$(\Delta\phi)$}.
\end{equation}
An object must satisfy all three of these criteria to be included in the 
sample.
The desired outcome of this analysis is the ability to predict the expected
contamination from spurious motions for a survey with parameters 
\hbox{$\mu_{lim}$}, \hbox{$(\Delta\mu)$}
and \hbox{$(\Delta\phi)$} given the `zero proper motion object' error distribution 
 described by equations~\ref{jdist} and~\ref{rdist}.  In order to
do this we consider the combined probability distribution $P(\hbox{$\mu_{b}$})P(\hbox{$\mu_{r}$})$
on the \hbox{$\mu_{b}$} -- \hbox{$\mu_{r}$}  plane.
\begin{figure}
\begin{center}
\setlength{\unitlength}{1mm}
\begin{picture}(70,50)
\includegraphics{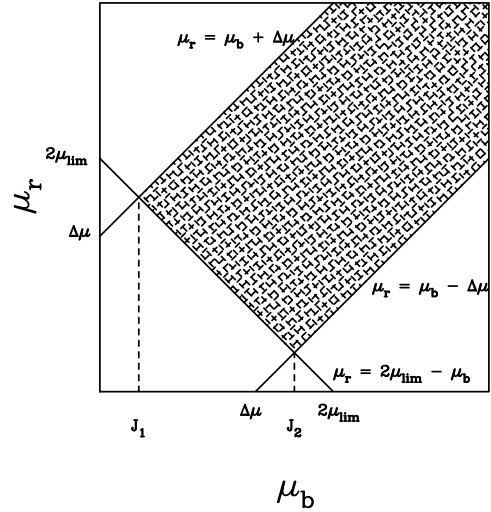} 
\end{picture}
\end{center}
\vspace{2cm}
 \caption{Objects will only be selected for the survey sample if their 
proper motions lie in the hatched region of the diagram.}
 \label{intdiag}
\end{figure}
The survey criteria described above effectively limit the selected sample 
to a specific area on the \hbox{$\mu_{b}$} -- \hbox{$\mu_{r}$}  plane, shown in Figure~\ref{intdiag}.
Given a realistic estimate of the (normalised) $P(\hbox{$\mu_{b}$})$ and $P(\hbox{$\mu_{r}$})$ 
distributions an integration over the hatched region is Figure~\ref{intdiag} yields
an estimate for the fraction of objects, X, belonging to the error distribution
likely to contaminate the sample. The integral
\begin{eqnarray}
\label{contamint}
\lefteqn{X = \int_{J_{1}}^{J_{2}} \int_{2\mu_{lim} - \mu_{b}}^{\mu_{b} + (\Delta\mu)}P(\mu_{b})P(\mu_{r})
{\rm d}\mu_{r} {\rm d}\mu_{b} \;  +} \nonumber  \\
& & \; \; \; \; \int_{J_{2}}^{\infty} \int_{2\mu_{b} - (\Delta\mu)}^{\mu_{b} + (\Delta\mu)}P(\mu_{b})P(\mu_{r})
{\rm d}\mu_{r} {\rm d}\mu_{b}
\end{eqnarray}
where
\begin{equation}
J_{1} = \frac{2\hbox{$\mu_{lim}$} - \hbox{$(\Delta\mu)$}}{2}
\end{equation}
and
\begin{equation}
J_{2} = \frac{2\hbox{$\mu_{lim}$} + \hbox{$(\Delta\mu)$}}{2}
\end{equation}
is easily calculated numerically for arbitrary \hbox{$\mu_{lim}$} and
\hbox{$(\Delta\mu)$}.  
Since this analysis concerns zero proper motion objects  whose
spurious motion arises purely from machine measurement error, we may assume
no preferred position angle, and the resulting fraction X will be cut by a further
$(2\hbox{$(\Delta\phi)$})/360$ (\hbox{$(\Delta\phi)$} in degrees) by the position angle selection criterion,
ie.
\begin{equation}
\label{pafrac}
X_{final} = \frac{X}{2(\Delta\phi)/360}.
\end{equation}
The calculation described above requires knowledge of the 
P(\hbox{$\mu_{b}$}) and P(\hbox{$\mu_{r}$}) `zero proper motion' error distributions. 
If equation~\ref{contamint} is to be used to calculate numbers of contaminants
it is important that $P(\hbox{$\mu_{b}$})$ and $P(\hbox{$\mu_{r}$})$ can be simultaneously rescaled
from normalised distributions.  The B and R data are therefore paired so that
 $P(\hbox{$\mu_{b}$})$ and $P(\hbox{$\mu_{r}$})$ contain the same objects and thus the same
number of objects.  For the
 purposes of this analysis this proper motion data is thought of as consisting
of two distinct distributions: the `zero proper motion object' error 
distribution, on which a distribution of real proper motions is superposed.
Any attempt to determine the error distribution from the measured distribution must
 use only the low proper motion data where the random errors of interest 
dominate systematics introduced by real proper motions.  In the example to 
be shown here the first 15 bins (1mas/yr bins) of the normalised 
proper motion distribution  is taken to be representative of the error distribution.

\begin{figure}
\begin{center}
\setlength{\unitlength}{1mm}
\begin{picture}(70,40)
\includegraphics{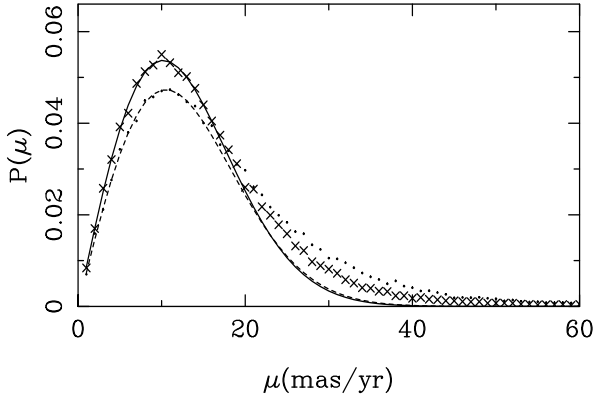} 
\end{picture}
\end{center}
\vspace{2cm}
 \caption{Normalised B (dashed line and dots) and R
 (solid line and crosses) measured 
data(points) and fitted error distributions (lines). This example uses the
first 15 data points to determine the error distribution.}
 \label{pmfit15}
\end{figure}

\begin{figure}
\begin{center}
\setlength{\unitlength}{1mm}
\begin{picture}(70,40)
\includegraphics{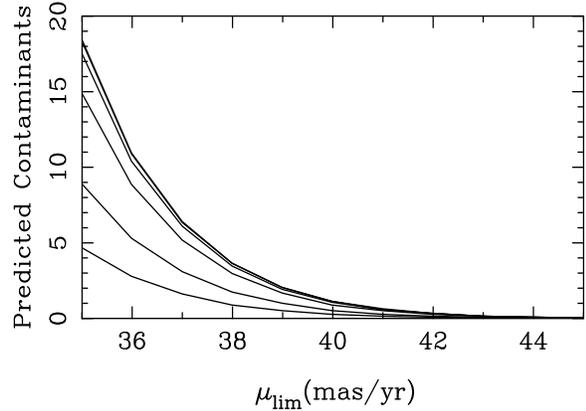} 
\end{picture}
\end{center}
\vspace{2cm}
 \caption{The results of numerical integration predicting the number of
contaminant objects in the survey sample as a function of survey proper motion
limit. The 
lower of the lines used a \hbox{$(\Delta\mu)$} of 5 mas/yr, with \hbox{$(\Delta\mu)$} rising to 50 mas/yr
for the top line (\hbox{$(\Delta\phi)$} was set to 90 throughout).  These calculations were
based on error distributions determined from the first 15 proper motion bins
of the measured \hbox{$\mu_{b}$} and \hbox{$\mu_{r}$} distributions.}
 \label{predcont15}
\end{figure}
The first 15 data points in both B and R are fitted with an assumed error distribution like
equation~\ref{rdist}.
These fits are shown along with the measured data in Figure~\ref{pmfit15}, with the
real data rising above the error fits representing real proper motions.
Since the error distributions exclude real motions they must be renormalised
before use in equation~\ref{contamint}.  The rescaling factor to be used
in calculating numbers of contaminants is taken to be the number
of objects contained in the error distributions.  This number will in general
be different for the B and R data, so the average is used.
Numerical methods can now be used to predict the number of contaminant
spurious proper motions for a range of \hbox{$\mu_{lim}$}, \hbox{$(\Delta\mu)$} and \hbox{$(\Delta\phi)$} using 
equations~\ref{contamint} and~\ref{pafrac}.  The result of a series of 
calculations is shown in Figure~\ref{predcont15}.
The predicted contamination falls rapidly below one object as
\hbox{$\mu_{lim}$} exceeds $\rm \sim42 mas/yr$.  Care must be taken however that the error
distributions  are not overly sensitive to the number of points in
the measured \hbox{$\mu_{b}$} and \hbox{$\mu_{r}$} distributions used in their calculation.  The error
distributions were therefore recalculated using the first 30 bins in the 
 measured \hbox{$\mu_{b}$} and \hbox{$\mu_{r}$} distribution and the 
predicted contamination plot redrawn using these new distributions .
The two calculations prediction of  the \hbox{$\mu_{lim}$} at which the survey
contamination drops to below 1 object are consistent to within $\rm \sim3 mas/yr$. 
Presented in Table~\ref{errlims} are the results of calculating the \hbox{$\mu_{lim}$}
at which the predicted number of contaminant objects falls below one for a 
range of magnitude cuts, \hbox{$(\Delta\mu)$} and derived $P(\hbox{$\mu_{b}$}),P(\hbox{$\mu_{r}$})$ distributions.
\begin{table*}
\begin{tabular}{|c|c|cccc|} \hline\hline
R magnitude  & No. of \hbox{$\mu_{b}$},\hbox{$\mu_{r}$} &  
\multicolumn{4}{c|} {\hbox{$\mu_{lim}$} (mas/yr) for varying \hbox{$(\Delta\mu)$}} \\ \cline{3-6}
cut          & data points used  & \hbox{$(\Delta\mu)$}=5 & \hbox{$(\Delta\mu)$}=10 & \hbox{$(\Delta\mu)$}=30
& \hbox{$(\Delta\mu)$}=50 \\ \hline
$R>20.5$ &  15 & 40 & 41 & 42 & 43 \\
        &  25 & 41 & 43 & 44 & 44 \\
$20.5>R>20$&  15 & 34 & 35 & 36 & 36 \\
        &  25 & 34 & 35 & 36 & 36 \\
$20>R>17$&  15 & 30 & 31 & 32 & 32 \\
        &  22 & 32 & 33 & 33 & 33 \\
$R<17$  &  15 & 31 & 32 & 33 & 33 \\
        &  25 & 34 & 35 & 36 & 36 \\
\hline\hline
\end{tabular}
 \caption{Calculated \hbox{$\mu_{lim}$} where predicted number of contaminants falls below
1 object.  Column two represents the number of bins in the measured \hbox{$\mu_{b}$} and
\hbox{$\mu_{r}$} distributions used to derive the 
$P(\hbox{$\mu_{b}$}),P(\hbox{$\mu_{r}$})$ distributions.}
 \label{errlims}
\end{table*}
Table~\ref{errlims} indicates that the \hbox{$(\Delta\mu)$} survey parameter has a
small bearing on the \hbox{$\mu_{lim}$} where $N_{contam} \sim 1$.  The
\hbox{$(\Delta\mu)$} criterion can therefore be relaxed substantially to ensure 
no real motions are rejected.  A similar argument is applicable to the
\hbox{$(\Delta\phi)$} parameter, where the small advantage gained by
 tightening the \hbox{$(\Delta\phi)$}
criterion again becomes increasingly outweighed by the potential for rejection
of real proper motions.  To conclude comment on this analysis, any
 \hbox{$\mu_{lim}$} greater than $\sim45$mas/yr
should eliminate contamination arising from normally distributed positional
 measurement errors, although 
thus far no account has been taken of spurious motions arising from other
 sources (eg. erroneous pairings).

\begin{figure}
\begin{center}
\setlength{\unitlength}{1mm}
\begin{picture}(70,40)
\includegraphics{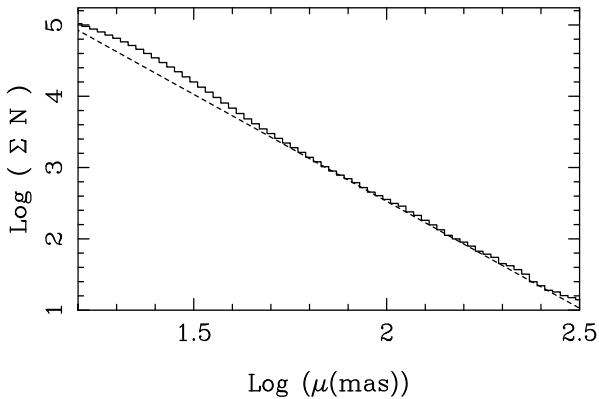} 
\end{picture}
\end{center}
\vspace{2cm}
 \caption{Cumulative number counts plot showing deviation from simple model
at $\mu \sim50 mas/yr$.}
 \label{culmplot}
\end{figure}

The second means of investigating the proper motion limit is via proper motion number counts. 
 Assuming the local Disc has constant stellar density and that Disc kinematics give rise to
an approximate inverse correlation between distance and proper motion, 
a plot of log cumulative number count (from large to small $\mu$) versus 
log $\mu$ should follow a straight line of gradient $-3$ (ie. $\log\sum N \propto -3 \log\mu$).
  In principle a proper motion limit for this survey could be 
obtained by determining the point at which our measured proper motions
deviate from this relation due to the existence of spurious motions.
In Figure~\ref{culmplot}
the cumulative number counts are plotted as a histogram.  If the
points between $\log \mu = 1.9$ and $\log \mu = 2.5$ are fit with a
straight line the resulting gradient is $(-2.998\pm0.071)$, in
excellent agreement with the idealised predicted slope of $-3$.
The fit is shown as a dashed line.
This finding compares favourably with a similar analysis of Luyten (1969,
1974, 1979) and 
Giclas (1971, 1978) common proper motion binary stars, which on an analogous plot
describe a straight line of significantly shallower gradient that the expected
$-3$ indicating increasing incompleteness with decreasing proper
motion (Oswalt \& Smith 1995).
Towards lower proper motions our data rises above the line, indicating
the onset of contamination at $\mu \sim50$ mas/yr,
lending credence to the findings of the previous analysis.

The final means of assessing the effects of varying \hbox{$\mu_{lim}$} is
the RPMD. The RPMD is, for all sample objects lacking follow up observations, the 
sole means of stellar population discrimination.  For this reason it is
worthwhile inspecting the RPMD of samples produced using various survey 
limits.  The principal concern is that the white dwarf locus be as distinct
as possible at all colours while maximising the sample size.
\begin{figure}
\begin{center}
\setlength{\unitlength}{1mm}
\begin{picture}(70,40)
\includegraphics{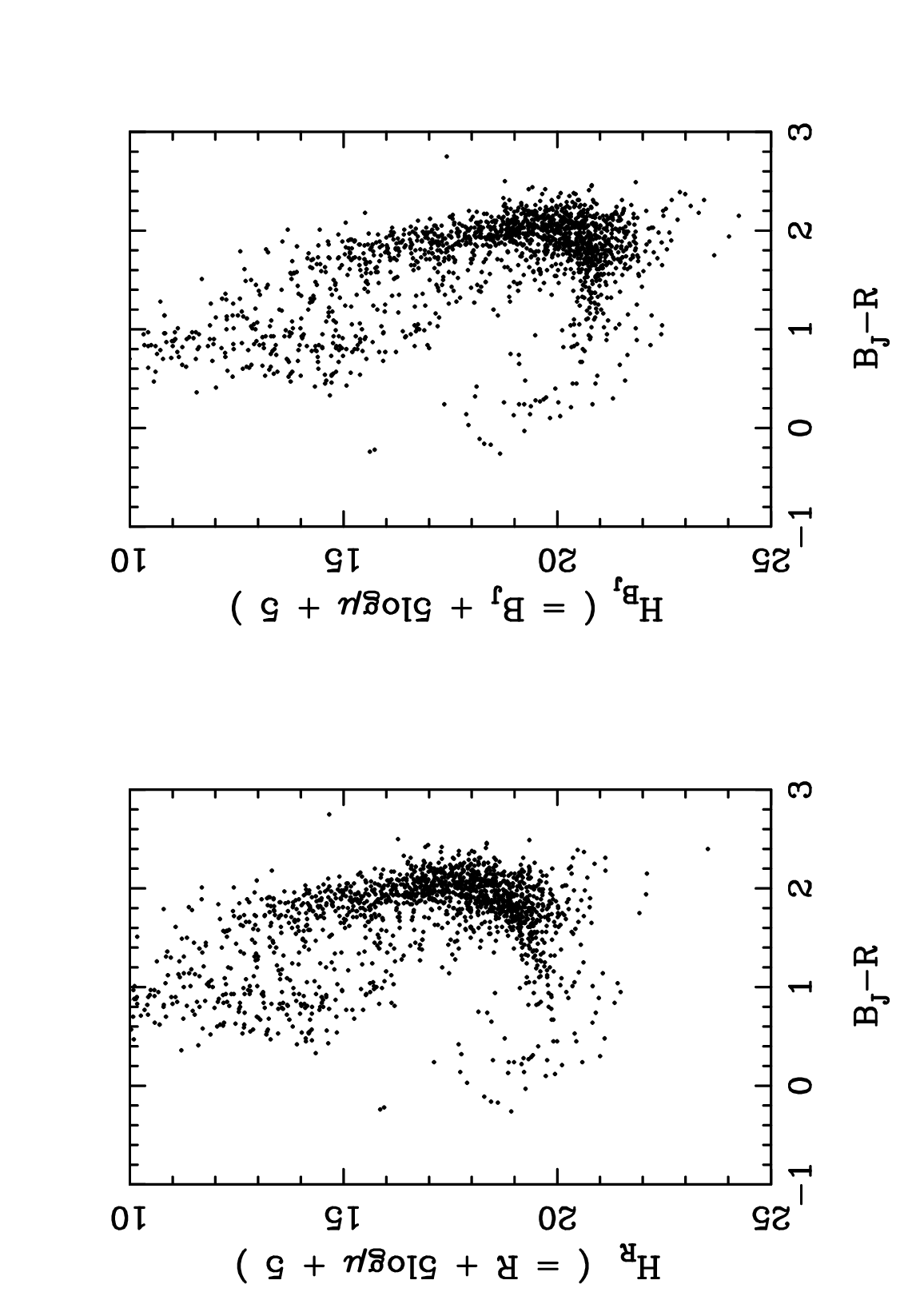} 
\end{picture}
\end{center}
\vspace{2cm}
 \caption{A RPMD of a sample with \hbox{$\mu_{lim}$}$= 50mas/yr$ for all objects.}
 \label{Hplt50mulim}
\end{figure}
\begin{figure}
\begin{center}
\setlength{\unitlength}{1mm}
\begin{picture}(70,40)
\includegraphics{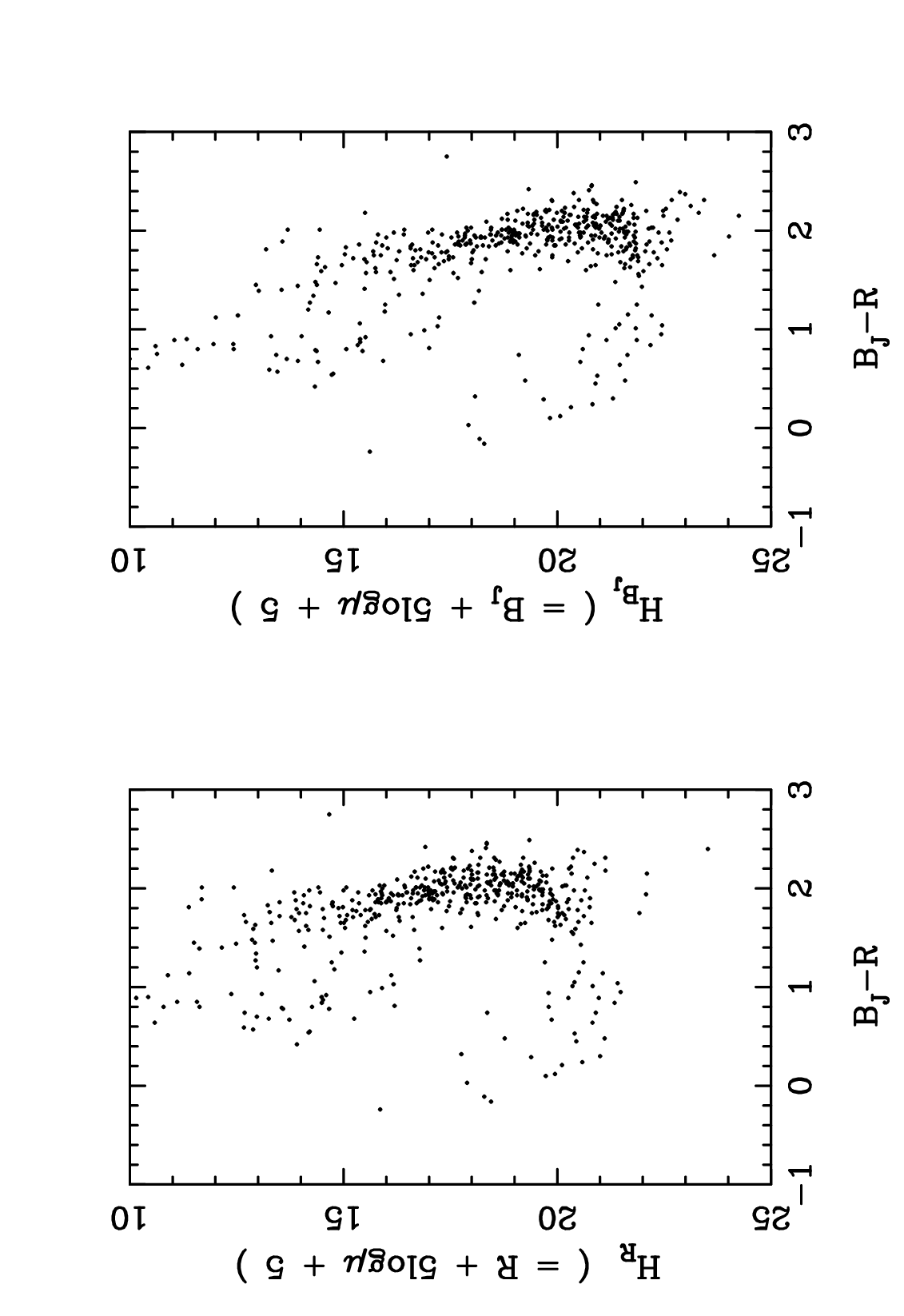} 
\end{picture}
\end{center}
\vspace{2cm}
 \caption{A RPMD of a sample with \hbox{$\mu_{lim}$}$= 80mas/yr$ for all objects.}
 \label{Hmulim80}
\end{figure}

It can be argued that there is little purpose in lowering \hbox{$\mu_{lim}$} for bright
 objects to anything near the level of potential contamination.
 Consider searching for an intrinsically faint star such  as a
 reasonably cool white dwarf, with an absolute R magnitude of $\sim14$,
 amongst survey objects with apparent R magnitudes as faint as $\sim19$.
 Such an object, having
a (conservative) tangential velocity of $\rm 40km/s$, would
have a proper motion of $\rm \sim80mas/yr$ -- well beyond the predicted 
contamination limit.  This argument becomes even more forceful 
for both intrinsically fainter and apparently brighter objects, implying
that a conservative \hbox{$\mu_{lim}$} is desirable for all but the faintest survey
objects.

\begin{figure}
\begin{center}
\setlength{\unitlength}{1mm}
\begin{picture}(70,40)
\includegraphics{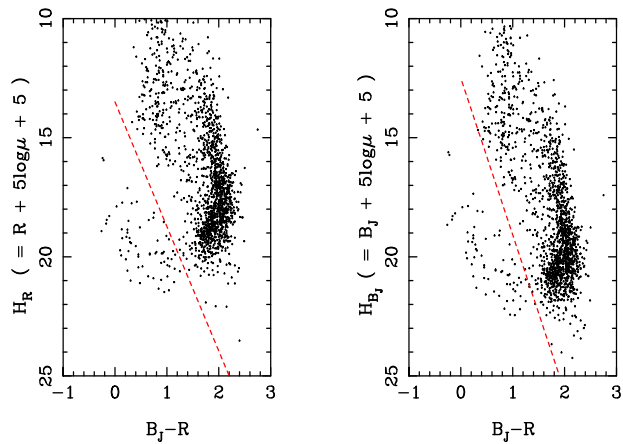} 
\end{picture}
\end{center}
\vspace{2cm}
 \caption{RPMD of the survey sample showing the white dwarf population 
divider used.}
 \label{samcut}
\end{figure}

Figure~\ref{Hplt50mulim} is a RPMD produced with a \hbox{$\mu_{lim}$} of 50 mas/yr
at every magnitude.
The main sequence and white dwarf loci are visible, as is a dense group of 
objects between the two at $H_{R}\sim19.5$.  This group of objects probably
consists of two sub-groups.  Firstly of course, the most likely sample contaminants
are faint objects creeping through just over the proper motion limit 
at this point.  Secondly, one would expect a large population  of
objects to lie on the RPMD where the $\mu$ and R
 distributions are most populated
, ie. at \hbox{$\mu_{lim}$} and at faint R (an object with $\mu = 50$ mas/yr and R=21
has $H_{\rm R} = 19.5$), and it should therefore not be automatically assumed 
that every object lying at this point in the RPMD is suspect.
  A cause for concern however, is the way this population bridges the 
gap between the main sequence and the white dwarf population, a property one
would expect from a contaminant locus rather than the detection of bona fide
proper motions.  If a more conservative limit of $\mu_{lim}$=60 mas/yr is 
adopted for objects with $R>20.5$ the RPMD (Figure~\ref{samcut}) looks
more promising, with the `contaminant locus' all but gone, leaving only the
expected mild confusion (Evans 1992)
 between populations at their faint extremity.
The adoption of an extremely conservative \hbox{$\mu_{lim}$} of 80 mas/yr for all objects
leads to an even more well defined RPMD (Figure~\ref{Hmulim80}), where the 
white dwarf population
discrimination is almost without exception unambiguous.

The survey parameters to be used to select the preliminary sample (ie. a
sample subject to further object-by-object scrutiny and potential rejection, with the
possibility of further objects being included the sample which lie just outside the
RPM cut on one or both RPMDs)
have been chosen with reference to the findings in this section.

Error analysis and number counts suggest a \hbox{$\mu_{lim}$} of 50 mas/yr,
 a \hbox{$(\Delta\mu)$} of
50 mas/yr and a \hbox{$(\Delta\phi)$} of 90 degrees should essentially eliminate contamination arising from
normally distributed measurement errors.  It was found, however, that the white
dwarf locus is insufficiently distinct in the RPMD obtained using these
survey parameters.  A slight restriction of \hbox{$\mu_{lim}$} for objects with 
$\rm R\geq20.5$
does much to solve the population discrimination problem. Therefore the
survey parameters given above, with the exception of a \hbox{$\mu_{lim}$} of 60 mas/yr 
for object with $\rm R\geq20.5$, appears to be an acceptable compromise between
sample maximisation and potential contamination and population discrimination
problems.

The RPM can be expressed in terms of tangential velocity ,$V_{T}$ ,and 
absolute magnitude, M :
\begin{equation}
H = M + 5\log_{10}V_{T} - 3.379.
\end{equation}
Evans (1992) has produced theoretical RPMDs by using  
expected $5\log_{10}V_{T}$ distributions and absolute magnitude - colour
relations for various populations.  Although these theoretical predictions were
made for specific fields and incorporated error estimates peculiar to that
work, they serve as a useful guide to the expected distribution of stellar populations
on the RPMD.
The RPMD white dwarf population locus is 
found to be an unambiguous population discriminator for all colours blue-wards
of $(O-E)\sim1.8$, with an increasing chance of contamination from the 
spheroid main sequence population red-wards of this colour.  
Transforming from
$(O-R)$ to $(B-R)$ (Evans, 1989 - equation 13), spheroid population contamination 
should become a problem red-wards of $(B-R)\sim1.6$ which in turn corresponds
to $(B_{J}-R)\sim1.2$. Indeed on inspection of Figure~\ref{samcut}, 
the white dwarf locus does begin to become confused at this colour.
This contamination
occurs solely from the direction of small RPM, and the RPMD can still be used
with some confidence as a population discriminator for objects with high
measured RPM red-wards of $(B_{J}-R)\sim1.2$.  In order to accommodate these
contamination considerations, the preliminary survey white dwarf sample is
 defined as those
objects blue-wards (in both plots) of the lines in Figure~\ref{samcut}.
The lines cut the top of the white dwarf locus at $(B_{\rm J}-R)\sim1.2$ but allow
slightly redder objects with higher RPMs into the sample.  The white dwarf
locus is unambiguous blue-wards of $(B_{\rm J}-R)\sim1.2$.

\begin{figure}
\begin{center}
\setlength{\unitlength}{1mm}
\begin{picture}(50,80)
\includegraphics{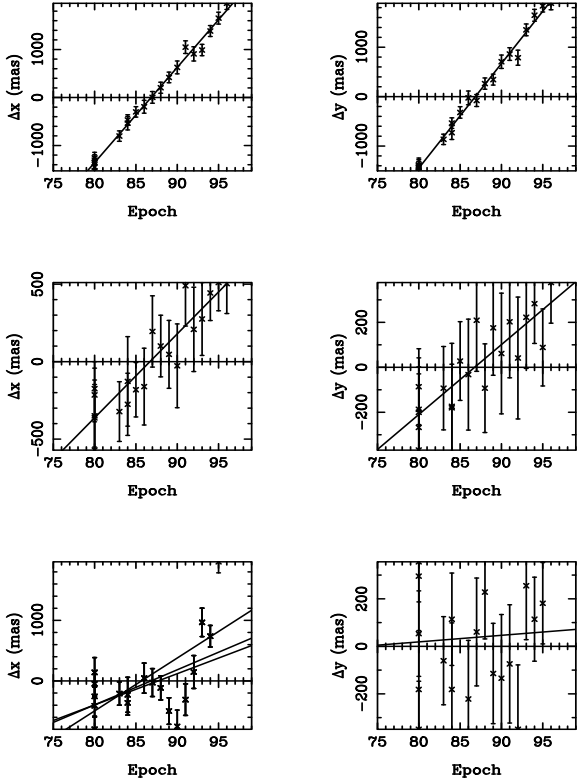} 
\end{picture}
\end{center}
\vspace{2cm}
 \caption{Some example proper motion plots from the R plate stacks.  The top plot
shows data for object KX27, the middle for KX18 and the lower for a rejected object}
 \label{Rexamples}
\end{figure}

\begin{figure}
\begin{center}
\setlength{\unitlength}{1mm}
\begin{picture}(50,80)
\includegraphics{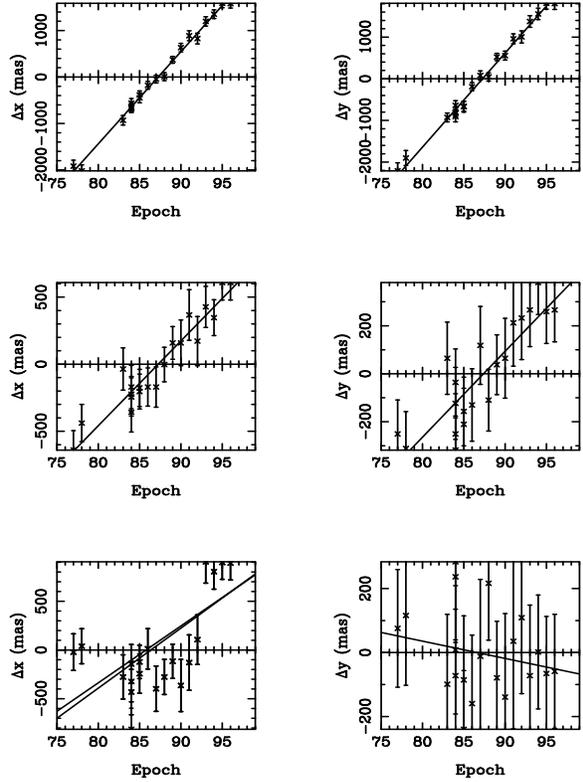} 
\end{picture}
\end{center}
\vspace{2cm}
 \caption{Proper motion plots complementary to Figure~\ref{Rexamples} from the 
$\rm B_{J}$ plate stack data}
 \label{Bexamples}
\end{figure}

Every object in the sample must appear in at least 15 stacks in each passband. 
The positional data as a function of time have been scrutinised for every
object selected as a white dwarf candidate, and those with dubious motions
rejected.  While such a process may seem rather arbitrary, it was necessary
to incorporate this screening stage in the sample extraction because simple
automated rejection algorithms such as the $\rm 3\sigma$ rejection routine
used here cannot be guaranteed to eliminate spurious motions.  Some
examples are shown in Figures~\ref{Rexamples} and~\ref{Bexamples}.
All three objects shown successfully satisfied all the survey criteria.  The
object plotted at the top of Figures~\ref{Rexamples} and~\ref{Bexamples} 
(KX27) shows a clear, genuine motion in both x and y in both passbands and 
was included in the final sample without hesitation.  The middle object (KX18)
has larger positional uncertainties and a smaller overall motion, but still
shows consistent, smooth motions and was also included.  The final object
shows evidence of large non-linear deviations in the last four epochs of the x
measures in both passbands. Although the bad-point rejection algorithm has removed
at least one datum from each x plot (as shown by the multiple straight line fits),
this object shows no evidence of proper motion based on the first 16 data points
and certainly cannot be considered a reliable proper motion object candidate.  
This object, along with 9 others, were rejected from the final WD sample.
These rejected objects tended either to have large offsets from a positional
distribution otherwise consistent with zero motion at either the first or last 
few epochs, as is the case with the rejected object described above; or the
positional measures had an unusually large scatter around the mean position, 
indicating the error in position was larger than the objects magnitude would 
suggest.  The reasons for these larger errors may be unusual image morphology
or the effect of proximity to a neighbouring object.
While this final rejection  procedure is unsatisfactory in
 terms of its lack of objectivity, it is certainly preferable
to inclusion of such objects in the final sample, or the introduction of extremely
stringent survey limits which would doubtless exclude genuinely interesting
objects.  The digitised images have also been inspected.

The final sample consists of 56 objects which fully satisfy the photometric,
proper motion  and RPM/colour survey limits.  A further two objects, which
satisfy the photometric and proper motion limits but fall marginally outside
the RPM/colour cut shown in Figure~\ref{samcut} have also been included 
after favourable follow up observations detailed in Section~\ref{followup}.

\section{Very high proper motion, faint object sensitivity limits}
\label{hipms}

In the previous Section, we discussed in some detail the checks made to
establish a clean astrometric and photometric catalogue. The availability of
plates over such a wide epoch range as detailed in Table~\ref{epochs} allowed us to
search for faint and/or very high proper motion objects in this field. This is
important for a number of reasons. For example, it is crucial to firmly
establish what the upper limit of detectable proper motion is for the 
methods used, and also to check if significant numbers of objects have been
missed because of this limit or because of the pairing algorithm used. Also,
it is important to check at fainter magnitudes for very dim, high proper 
motion objects since it is the coolest (and therefore faintest) objects that
constrain the age determination based on the turn--over in the WDLF. It is
also interesting to search for very faint, high proper motion halo WDs in the
light of the current debate concerning the origin of the dark lensing bodies
detected in microlensing experiments (eg.~Isern et al.~1998 and references
therein).

We performed three experiments to investigate high proper motion and/or faint
objects:
\begin{enumerate}
\item Within the restricted epoch range 1992 to~1996, ie.~five separate epochs,
each consisting of a stack of four R~plates, we used completely independent
software employing a `multiple--pass' pairing technique aimed specifically at
detecting high proper motion objects. This software has successfully detected
a very cool, high proper motion degenerate WD~0346+246 elsewhere (Hambly,
Smartt \& Hodgkin~1997). The pairing algorithm described previously used a
$200\mu$m search radius over 19~yr resulting in an upper limit of 
$\sim 1$~arcsec~yr$^{-1}$, whereas in the multiple--pass test we used a maximum 
search radius of $650\mu$m over a 4~yr baseline, theoretically enabling 
detection of objects with annual motions as high as $\sim10$~arcsec. The 
highest proper motion object detected in the catalogue 
was relatively bright (${\rm R}\sim14$), with 
$\mu\sim0.8$~arcsec~yr$^{-1}$. This experiment revealed two 
$\mu\sim0.8$~arcsec~yr$^{-1}$ objects
, the one mentioned above and another slightly fainter object (${\rm R}\sim15$)
; no objects were found with motions larger than this. The colours and reduced
proper motions of the two objects indicate that they are M-type dwarfs.  The
second object was undetected in the catalogue 
due to a spurious pairing, an increasingly likely
scenario for high proper motion objects detected without a multiple pass algorithm
since they move substantially from their master frame position.
 We note
that all the expected objects having $\mu>0.2$~arcsec~yr$^{-1}$ detected in the
catalogue were also found by this procedure.
\item Using the procedure described in the previous Section, but with a relaxed
minimum number of epochs, high proper motion objects were searched for. With a
$200\mu$m pairing requirement over a maximum epoch separation of 7~yr, the
upper limit of proper motion was $\sim 1.9$~arcsec~yr$^{-1}$.
 The two objects having
$\mu\sim0.8$~arcsec~yr$^{-1}$ found in the previous experiment were also 
recovered here; again, no objects were found with motions larger than this.
\item To investigate the possibility of fainter objects, we stacked up the 
R~band material in groups of 16~plates at epochs~1980, 1983 to~1986,
1987 to~1991 and~1992 to~1996. Obviously, over any individual four year period
an object having a proper motion greater than $\sim1$~arcsec~yr$^{-1}$ will
have an extended image and will not be detected to the same level of faintness
as a stationary star; nonetheless, the $\sim0.75^{\rm m}$ increase in depth
afforded by going from 4~to 16~plate stacks (eg.~Knox et al.~1998) at least
allows us to investigate the possible existence of objects having
$\mu\sim0.5$~arcsec~yr$^{-1}$ down to ${\rm R}\sim23$ (100\% complete to
${\rm R}\sim22$) over an area of 25~square degrees. In this experiment, all the
objects expected from the catalogue were recovered; in addition, one star was
found having ${\rm R}\sim20$, $\mu=0.47$~arcsec~yr$^{-1}$ at 
${\rm PA}=179^{\circ}$ and RA,DEC~=~21h30m8.553s, $-44^{\circ}$46'24.09
(J2000.0). This object is the M--type dwarf `M20' discovered in the
photometric survey of Hawkins \& Bessell~(1988) and has 
${\rm B}_{\rm J}\sim23$. The faintness in the blue passband is the reason that
this object is absent from the catalogue. Once more, no other high proper 
motion, fainter stars were found.
\end{enumerate}
These three experiments allow us to be confident that there is no large
population of objects having $\mu\ga1$~arcsec~yr$^{-1}$ down to faintness
limits of ${\rm R}\sim22$ and ${\rm B}_{\rm J}\sim23$. Furthermore, the
cut--off in the WD sequence seen in the reduced proper motion diagrams is
real, and not an artefact of incompleteness.

\section{Follow up Observations}
\label{followup}

While the RPMD technique is a powerful population discriminator, it is 
desirable to obtain follow up observations of a sub-set of sample members.
The principle motivation for this is to explicitly demonstrate the applicability
of our survey technique by confirming the WD status of the sample objects
via spectroscopy.  Spectroscopic observations red-wards of the WD cut in
 the RPMD may also be used to investigate the possibility of ultra-cool
white dwarfs existing in the sample; and as a corollary to this, such 
observations allow clearer population delineation in the RPMD.
In addition, photometric observations through standard 
filters ensure confidence in photographic-to-standard photometry 
transformations, provide useful independent checks of stellar parameters
(eg $T_{\rm eff}$) derived from fits to photometry and may in future allow alternative
 population discrimination via eg. colour-colour plots.

\subsection{Spectroscopy}

This project was allocated 3 nights of observing time in 1996 between the
8th and 10th of August, and a further 3 nights in 1997 between the 5th and
7th of August on the 3.9m Anglo-Australian Telescope.  A spectral coverage
of 4000-7500$\rm\AA$ was obtained with the RGO spectrograph and 300B grating
in conjunction with the Tek CCD.  Various standards were observed throughout
each usable night and CuAr lamp exposures used for wavelength calibration.
The data were reduced following standard procedures within the IRAF
\footnote{IRAF is distributed by the National Optical Astronomy 
Observatories, which is operated by the Association of Universities for 
Research in Astronomy Inc., under contract with the National Science 
Foundation of the United States of America.}
environment.
\begin{figure}
\begin{center}
\setlength{\unitlength}{1mm}
\begin{picture}(70,50)
\includegraphics{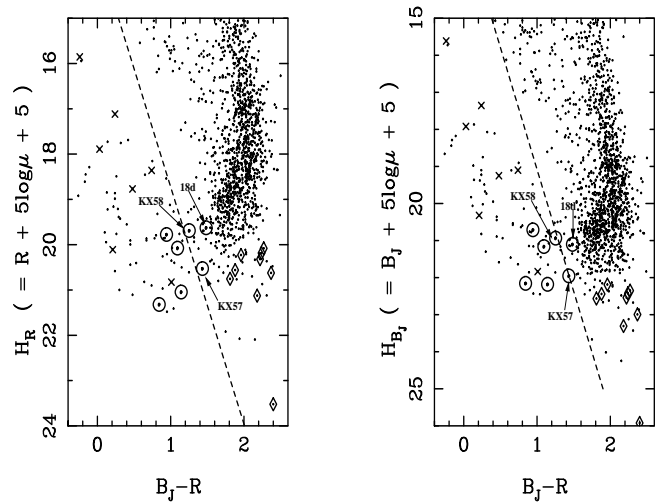} 
\end{picture}
\end{center}
\vspace{2cm}
 \caption{RPMD of the survey sample with spectroscopically observed objects
numbered and 
highlighted.  Diamonds denote objects identified as subdwarfs or M dwarfs,
crosses an immediate firm WD identification.  Dotted circles represent objects
with ambiguous spectra (see text).The labelled objects have their spectra 
discussed in detail below.}	
 \label{spechplot}
\end{figure}

The strategy behind these observations was to
define as clearly as possible the constituents of the lower portion of the
RPMD, from the blue end of the WD locus down to the extremely high H objects
below the M stars.  The spectroscopically observed
proper motion sample objects are shown in Figure~\ref{spechplot}.  Both the
 unambiguous bluer region
of the WD locus and the portion of RPMD lying red-ward of our WD cut generally
consist of fairly bright objects.  These regions have been probed 
spectroscopically, albeit more sparsely than the cool WD region consisting
of mostly very faint objects.  A combination of poor weather conditions
at the AAT and the faintness of our CWD sample has rendered high
 signal-to-noise spectra of these objects unobtainable thus far.

The bluest objects have clearly defined Hydrogen Balmer lines with equivalent
widths equal to or in excess of those typical for DA white dwarfs of 
similar colour (Greenstein and Liebert, 1990).  The objects red-wards
of our sample cut below the main sequence (MS) show spectra clearly
distinct from 
cool WDs.  Both subdwarfs and high velocity M dwarfs have been identified
in this region of the RPMD.

A star lying near the cut-off region of the RPMD clearly showing 
 the absence of strong metal 
features that would be present even in a low-metallicity subdwarf 
 is very likely a CWD.  We apply this
line of argument to our data by selecting from
 the recent models of Hauschildt et al. (1998) a spectrum appropriate to a
subdwarf (metallicity $\rm[M/H]\geq-2$) of similar effective temperature to a 
given CWD sample object spectrum.
The model spectrum is smoothed to the approximate resolution of the AAT spectra
and multiplied through by a synthetic noise spectrum commensurate with the AAT
CWD spectrum in question.  The only features likely to be visible after this
procedure are $\sim5200\rm\AA$ MgI and the $\sim4300\rm\AA$ CH G-band, and
it is in these regions that we look for any evidence that our CWD sample objects
are in fact subdwarfs.  

This procedure has been undertaken for all CWD candidate spectra.  Of these, the 
spectra of the three objects lying within the region of serious potential 
subdwarf contamination
are displayed, and we restrict comment on the investigation of the 
bluer objects to the statement that none show any evidence of subdwarf like
spectral features.
\begin{figure}
\begin{center}
\setlength{\unitlength}{1mm}
\begin{picture}(50,90)
\includegraphics{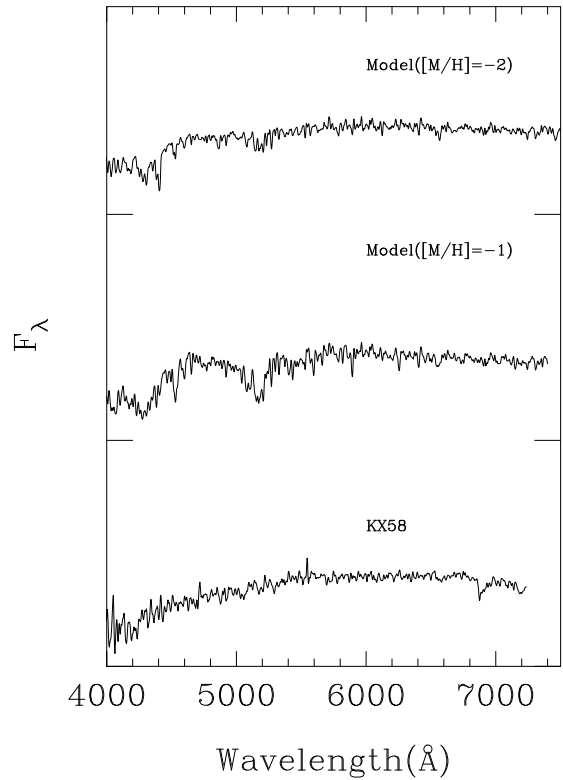} 
\end{picture}
\end{center}
\vspace{2cm}
 \caption{Spectrum of KX58 with comparison model subdwarf spectra}
 \label{KX58specplot}
\end{figure}
KX58, displayed in Figure~\ref{KX58specplot}, is a convincing CWD candidate, showing
 a smooth continuum spectrum with no
suggestion of the metal features apparent in the model subdwarf spectra.
\begin{figure}
\begin{center}
\setlength{\unitlength}{1mm}
\begin{picture}(50,90)
\includegraphics{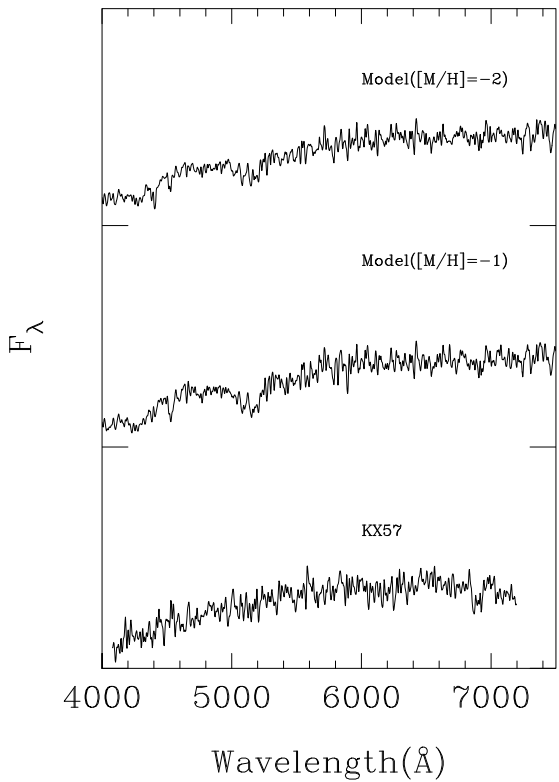} 
\end{picture}
\end{center}
\vspace{2cm}
 \caption{Spectrum of KX57 with comparison model subdwarf spectra}
 \label{KX57specplot}
\end{figure}
The noisier spectrum of KX57 in Figure~\ref{KX57specplot} also shows no evidence
of subdwarf features, although poorer signal-to-noise makes the identification
less certain.
\begin{figure}
\begin{center}
\setlength{\unitlength}{1mm}
\begin{picture}(50,90)
\includegraphics{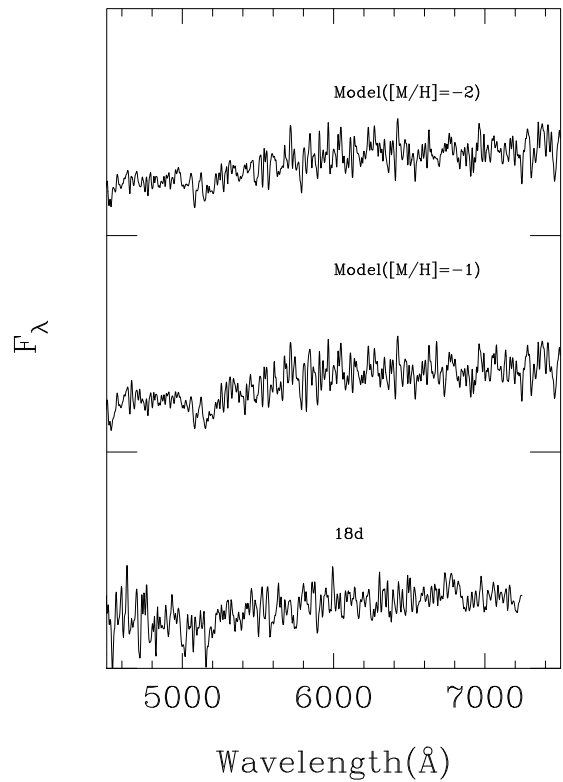} 
\end{picture}
\end{center}
\vspace{2cm}
 \caption{Spectrum of 18d with comparison model subdwarf spectra}
 \label{18dspecplot}
\end{figure}
The object 18d lies significantly red-ward of our CWD cut-off on the RPMD,  and 
unfortunately its spectrum (Figure~\ref{18dspecplot}) is extremely noisy.
While it is difficult to draw any conclusions from such poor data, the 
dip at $\sim5200\rm\AA$ is a reasonable indication that this object is
 a subdwarf or MS star.  
Thus the position of 18d on the RPMD in conjunction with  
the spectral data evidence means this object does not warrant inclusion in our CWD
sample.

To summarise the findings of our spectroscopic survey, the only objects
 showing notable deviation from expected WD spectra are the objects below
the M dwarf portion of the RPMD (diamonds in Figure~\ref{spechplot}) and 
the more ambiguous case of object 18d discussed above.

\subsection{Photometry}
\label{photom}

CCD photometry of a subsample of our CWD sample was obtained between 29th
of July and the 4th of August 1997 on the 1m telescope of the South African
Astronomical Observatory in Sutherland.  Johnson-Cousins V, R, I photometry
 was obtained for all program stars on the Tek (512x512) CCD, with B measures
also acquired for sufficiently bright objects.  E-region standards 
were observed continuously through each usable night.  Observed
magnitudes with associated errors are displayed in Table~\ref{saaophot}.

\begin{table}
\begin{tabular}{|c|cccc|} \hline\hline
Object & B & V & R & I \\
       & $\sigma_{\rm B}$  & $\sigma_{\rm V}$  & $\sigma_{\rm R}$ 
 & $\sigma_{\rm I}$\\ \hline
KX14(1)	&  --	& 21.43 & 20.78	& 20.17 \\
	&	& 0.04	& 0.09	& 0.07 	\\
KX14(2)	&  --	& 21.47	& 20.75	& 20.20 \\
	&	& 0.04	& 0.07	& 0.07	\\
KX15	& 19.10	& 18.53	& 18.08 & 17.69 \\
	& 0.03	& 0.02	&  0.03	&  0.03	\\
KX20	& 16.14	& 16.12	& 16.16	& 16.22	\\
	& 0.02	& 0.02	& 0.02	& 0.02	\\
KX22	&  --	& 19.94	& 19.66 & 19.55 \\
	&  	& 0.03	& 0.04	& 0.04  \\
KX23	& 19.52	& 19.05	& 18.74	& 18.61	\\
	& 0.03	& 0.01	& 0.01	& 0.03  \\
KX27	&  --	& 18.15	& 17.97	& 17.91	\\
	& 	& 0.02	& 0.02	& 0.03 	\\
KX29	& 18.67	& 18.36	& 18.17	& 17.90 \\
	& 0.03	& 0.02	& 0.03	& 0.03	\\
KX30	&  --	& 21.12	& 20.64	& 20.19 \\
	&	& 0.03	& 0.05 	&  0.08	\\
KX33	&  --	& 21.22 & 20.69	& 20.10 \\
	&	& 0.04	& 0.06	&  0.07	\\
KX39	&  --	& 21.19 & 20.66	& 20.18	\\
	& 	& 0.04  & 0.06	& 0.06  \\
KX41a	& 20.26	& 19.38	& 18.93	& 18.41	\\
	& 0.05	& 0.03	& 0.04	& 0.04	\\
KX41b	& 20.28	& 19.33	& 18.81	& 18.36 \\
	& 0.04	& 0.03	& 0.04 	& 0.04	\\
KX44	&  -- 	& 21.37	& 20.81	& 20.42	\\
	&	& 0.05	& 0.07	& 0.08	\\
KX53	&  --	& 20.96	& 20.70	& 20.21 \\
	&	& 0.03	& 0.05 	& 0.06	\\
KX57	&  --	& 21.09	& 20.40	& 19.84	\\
	&	& 0.04	& 0.06	& 0.08	\\
\hline\hline
\end{tabular}
 \caption{Johnson-Cousins CCD photometry taken at SAAO for selected members 
of our CWD sample. Object KX41 was resolved as a double-degenerate
 on the CCD frame and thus has photometry for each component. Object KX14
had two independent sets of observations, shown as (1) and (2)}
 \label{saaophot}
\end{table}
These observed magnitudes provide an independent check on the accuracy
of the SuperCOSMOS photographic photometry, and we use the deviations
of $\rm m_{photographic}$ from $\rm m_{CCD}$ to obtain errors on the B, V,
R, I photographic photometry of 0.17, 0.14, 0.13 and 0.16 respectively.
The CCD photometry also allowed tighter estimates of effective temperature
to be derived for observed objects, although they did not provide the hoped
for useful constraints on log g.

\section{Sample Analysis}
\label{samanal}

Previous studies of CWD samples (eg LDM, BRL) have often benefited from a
comprehensive and wide ranging observational database, including high quality 
spectra, optical and IR photometry and parallaxes.  These observations,
 in conjunction with detailed WD models, allow determinations of stellar
 parameters such as effective temperature, log g, atmospheric composition,
mass and bolometric luminosity.  
However, since this is a relatively
new project and is concerned with stars of unusually faint apparent magnitude,
such a database does not yet exist for this sample.  It is therefore necessary
for us to restrict our analysis, in the first place by exploiting the
homogeneity of WD masses by assuming a common typical log g for our entire
CWD sample (the 60 stars with 
measured log g in BRL have a mean surface gravity $\overline{\log{g}} =
 8.099\pm0.044$), and secondly by treating the atmospheric constituent of each
star as an unknown parameter whose influence on the resulting WDLF must
be determined later.

Bergeron et al. (1995) have published a detailed grid of model predictions
for Johnson-Cousins U, B, V, R, I (and IR) photometry and bolometric corrections
as a function of effective temperature and log g.
Making our assumption
that log g is always equal to 8, values for effective 
temperature and bolometric luminosity assuming both a H and He atmosphere
can be calculated for every sample object.

Fitting for $T_{\rm eff}$ is achieved by interpolating the model grid
at 10K intervals for the colour indices $(U-B)$, $(B-R)$, $(V-R)$ and $(V-I)$.
We then evaluate $\chi^{2}$ at each $T_{\rm eff}$ interval using all available
colour indices for the object in question.  While the photographic photometry
is effective in adequately constraining $T_{\rm eff}$, smaller errors
 are obtainable with the SAAO CCD photometry, and it is used where
available.  The resulting
$\chi^{2}$, $T_{\rm eff}$ distribution yields a fitted value for $T_{\rm eff}$
and an estimate of the associated error, which can be used to read off
interpolated values of absolute V magnitude and 
bolometric luminosity.  This procedure is performed
for each object for both an assumed H and He atmosphere.
\begin{table*}
\begin{tabular}{|c|cccc|cccc|} \hline\hline
Atmosphere: & \multicolumn{4}{c|} H  & \multicolumn{4}{c|} {He} \\ \hline
	& $T_{\rm eff}$ & $M_{\rm bol}$ & $V_{\rm tan}$ & $V_{\rm max}^{-1}$ &
 $T_{\rm eff}$ & $M_{\rm bol}$ & $V_{\rm tan}$ & $V_{\rm max}^{-1}$ \\ 
Object	& $\rm (K^{+1\sigma}_{-1\sigma})$ & & km/s & $\rm (10^{-4} pc^{-3})$ & $\rm (K^{+1\sigma}_{-1\sigma})$ & & km/s &
 $\rm (10^{-4} pc^{-3})$ \\ \hline
KX 1  &  $ 4700^{ 5090}_{ 4320}$  &  $15.14^{14.79}_{15.51}$  &   76.0  &  0.956  &  $ 5250^{ 5520}_{ 5030}$  &  $14.68^{14.45}_{14.86}$  &  102.2  &  0.436  \\
KX 2  &  $ 5980^{ 6620}_{ 5470}$  &  $14.08^{13.63}_{14.47}$  &   72.5  &  0.209  &  $ 6340^{ 6950}_{ 5890}$  &  $13.85^{13.45}_{14.18}$  &   84.4  &  0.144  \\
KX 3  &  $ 5800^{ 6340}_{ 5340}$  &  $14.22^{13.82}_{14.58}$  &   73.7  &  0.317  &  $ 6050^{ 6540}_{ 5710}$  &  $14.06^{13.72}_{14.31}$  &   83.5  &  0.231  \\
KX 4  &  $ 5060^{ 5480}_{ 4680}$  &  $14.81^{14.46}_{15.16}$  &   83.7  &  0.546  &  $ 5460^{ 5780}_{ 5210}$  &  $14.50^{14.26}_{14.71}$  &  103.5  &  0.315  \\
KX 5  &  $ 7590^{ 8800}_{ 6800}$  &  $13.03^{12.38}_{13.51}$  &   71.2  &  0.218  &  $ 7780^{ 8820}_{ 6990}$  &  $12.96^{12.41}_{13.43}$  &   76.2  &  0.185  \\
KX 6  &  $ 5000^{ 5420}_{ 4620}$  &  $14.86^{14.51}_{15.21}$  &   47.2  &  1.026  &  $ 5460^{ 5790}_{ 5210}$  &  $14.50^{14.25}_{14.71}$  &   60.2  &  0.536  \\
KX 7  &  $ 7360^{ 8630}_{ 6540}$  &  $13.16^{12.47}_{13.68}$  &   99.6  &  0.098  &  $ 7400^{ 8320}_{ 6700}$  &  $13.18^{12.67}_{13.61}$  &  103.3  &  0.090  \\
KX 8  &  $ 8670^{10450}_{ 7280}$  &  $12.45^{11.63}_{13.21}$  &   98.9  &  0.100  &  $ 8120^{ 9290}_{ 7240}$  &  $12.77^{12.19}_{13.27}$  &   89.6  &  0.125  \\
KX 9  &  $ 5300^{ 5750}_{ 4930}$  &  $14.61^{14.25}_{14.93}$  &   87.4  &  0.427  &  $ 5670^{ 5990}_{ 5390}$  &  $14.34^{14.10}_{14.56}$  &  105.2  &  0.266  \\
KX10  &  $ 4890^{ 5260}_{ 4530}$  &  $14.96^{14.64}_{15.30}$  &   57.4  &  0.738  &  $ 5300^{ 5560}_{ 5080}$  &  $14.63^{14.43}_{14.82}$  &   71.5  &  0.415  \\
KX11  &  $ 3970^{ 4370}_{ 3750}$  &  $15.87^{15.45}_{16.12}$  &   42.2  &  2.892  &  $ 4720^{ 4940}_{ 4530}$  &  $15.14^{14.94}_{15.32}$  &   58.8  &  1.156  \\
KX12  &  $ 4600^{ 4970}_{ 4240}$  &  $15.23^{14.89}_{15.59}$  &   22.5  &  4.823  &  $ 5290^{ 5550}_{ 5050}$  &  $14.64^{14.43}_{14.85}$  &   32.6  &  1.707  \\
KX13  &  $ 6370^{ 7010}_{ 5850}$  &  $13.80^{13.38}_{14.18}$  &  149.7  &  0.168  &  $ 6540^{ 7200}_{ 6090}$  &  $13.72^{13.30}_{14.03}$  &  162.8  &  0.138  \\
KX14  &  $ 3900^{ 4380}_{ 3750}$  &  $15.95^{15.44}_{16.12}$  &  102.9  &  3.147  &  $ 4600^{ 4780}_{ 4440}$  &  $15.48^{15.30}_{15.64}$  &  123.7  &  1.883  \\
KX15  &  $ 5560^{ 5710}_{ 5430}$  &  $14.40^{14.28}_{14.50}$  &   35.7  &  1.330  &  $ 5750^{ 5860}_{ 5650}$  &  $14.55^{14.50}_{14.59}$  &   32.6  &  1.716  \\
KX16  &  $ 5800^{ 6320}_{ 5360}$  &  $14.22^{13.84}_{14.56}$  &   70.3  &  0.243  &  $ 6070^{ 6550}_{ 5730}$  &  $14.04^{13.71}_{14.30}$  &   80.2  &  0.175  \\
KX17  &  $ 8300^{10100}_{ 7140}$  &  $12.64^{11.77}_{13.30}$  &   48.3  &  0.593  &  $ 7960^{ 9090}_{ 7130}$  &  $12.86^{12.28}_{13.34}$  &   45.8  &  0.682  \\
KX18  &  $ 7690^{ 9050}_{ 6830}$  &  $12.97^{12.26}_{13.49}$  &   84.6  &  0.144  &  $ 7740^{ 8760}_{ 6950}$  &  $12.98^{12.44}_{13.45}$  &   87.7  &  0.132  \\
KX19  &  $15620^{21150}_{ 9690}$  &  $ 9.86^{ 8.54}_{11.96}$  &  166.7  &  0.033  &  $11830^{15400}_{ 9850}$  &  $11.13^{ 9.97}_{11.93}$  &  126.8  &  0.058  \\
KX20  &  $15380^{16440}_{14450}$  &  $ 9.93^{ 9.63}_{10.20}$  &   38.2  &  1.118  &  $14660^{15990}_{13740}$  &  $11.08^{10.93}_{11.25}$  &   32.0  &  1.802  \\
KX21  &  $ 6120^{ 6710}_{ 5640}$  &  $13.98^{13.57}_{14.34}$  &  104.2  &  0.201  &  $ 6360^{ 6910}_{ 5930}$  &  $13.84^{13.48}_{14.15}$  &  116.3  &  0.154  \\
KX22  &  $ 6970^{ 7350}_{ 6690}$  &  $13.40^{13.17}_{13.58}$  &   76.9  &  0.181  &  $ 7040^{ 7380}_{ 6760}$  &  $13.68^{13.46}_{13.88}$  &   68.0  &  0.246  \\
KX23  &  $ 6570^{ 6750}_{ 6410}$  &  $13.66^{13.55}_{13.77}$  &   57.0  &  0.384  &  $ 6640^{ 6820}_{ 6470}$  &  $13.96^{13.84}_{14.08}$  &   49.5  &  0.556  \\
KX24  &  $ 4950^{ 5340}_{ 4600}$  &  $14.91^{14.58}_{15.23}$  &   21.4  &  5.556  &  $ 5370^{ 5660}_{ 5140}$  &  $14.58^{14.35}_{14.77}$  &   26.8  &  2.960  \\
KX25  &  $ 9210^{11210}_{ 7860}$  &  $12.18^{11.32}_{12.88}$  &  155.8  &  0.044  &  $ 8890^{10340}_{ 7830}$  &  $12.38^{11.72}_{12.93}$  &  146.9  &  0.050  \\
KX26  &  $ 5020^{ 5420}_{ 4660}$  &  $14.85^{14.51}_{15.17}$  &   51.6  &  0.568  &  $ 5420^{ 5720}_{ 5190}$  &  $14.54^{14.30}_{14.73}$  &   63.8  &  0.328  \\
KX27  &  $ 7340^{ 7640}_{ 7050}$  &  $13.18^{13.00}_{13.35}$  &  117.2  &  0.107  &  $ 7230^{ 7550}_{ 6940}$  &  $13.56^{13.35}_{13.75}$  &   99.4  &  0.158  \\
KX28  &  $ 8660^{10440}_{ 7340}$  &  $12.45^{11.63}_{13.18}$  &  248.4  &  0.053  &  $ 8220^{ 9410}_{ 7320}$  &  $12.72^{12.13}_{13.23}$  &  230.0  &  0.063  \\
KX29  &  $ 7850^{ 8220}_{ 7520}$  &  $12.88^{12.68}_{13.07}$  &   33.8  &  1.547  &  $ 7920^{ 8270}_{ 7610}$  &  $13.13^{12.93}_{13.32}$  &   30.7  &  2.023  \\
KX30  &  $ 5150^{ 5490}_{ 4840}$  &  $14.74^{14.45}_{15.01}$  &  119.3  &  0.508  &  $ 5280^{ 5540}_{ 5070}$  &  $14.84^{14.63}_{15.03}$  &  109.8  &  0.632  \\
KX31  &  $ 6640^{ 7430}_{ 6060}$  &  $13.62^{13.12}_{14.02}$  &   43.7  &  0.775  &  $ 6780^{ 7500}_{ 6250}$  &  $13.56^{13.12}_{13.92}$  &   46.8  &  0.643  \\
KX32  &  $13060^{16410}_{10400}$  &  $10.65^{ 9.64}_{11.65}$  &   35.9  &  1.323  &  $10810^{13510}_{ 9190}$  &  $11.52^{10.55}_{12.23}$  &   28.7  &  2.437  \\
KX33  &  $ 4650^{ 4980}_{ 4320}$  &  $15.18^{14.88}_{15.51}$  &   81.4  &  1.095  &  $ 4940^{ 5160}_{ 4760}$  &  $15.15^{14.95}_{15.32}$  &   80.7  &  1.119  \\
KX34  &  $ 6940^{ 7860}_{ 6310}$  &  $13.42^{12.88}_{13.84}$  &   36.9  &  1.222  &  $ 7080^{ 7920}_{ 6460}$  &  $13.37^{12.88}_{13.77}$  &   39.4  &  1.019  \\
KX35  &  $ 8390^{10240}_{ 7240}$  &  $12.59^{11.72}_{13.24}$  &   49.1  &  0.566  &  $ 8090^{ 9240}_{ 7230}$  &  $12.79^{12.21}_{13.28}$  &   46.9  &  0.639  \\
KX36  &  $ 7270^{ 8260}_{ 6570}$  &  $13.22^{12.66}_{13.66}$  &  143.8  &  0.109  &  $ 7480^{ 8430}_{ 6780}$  &  $13.13^{12.61}_{13.56}$  &  155.8  &  0.091  \\
KX37  &  $20560^{27560}_{15870}$  &  $ 8.65^{ 7.36}_{ 9.79}$  &  169.5  &  0.032  &  $14720^{26300}_{11470}$  &  $10.17^{ 7.61}_{11.27}$  &  129.3  &  0.055  \\
KX38  &  $10250^{13600}_{ 8620}$  &  $11.71^{10.47}_{12.47}$  &   81.9  &  0.155  &  $ 9910^{11890}_{ 8580}$  &  $11.90^{11.11}_{12.53}$  &   75.9  &  0.187  \\
KX39  &  $ 4910^{ 5230}_{ 4600}$  &  $14.95^{14.67}_{15.23}$  &   62.4  &  0.760  &  $ 5090^{ 5320}_{ 4910}$  &  $15.01^{14.81}_{15.17}$  &   58.5  &  0.903  \\
KX40  &  $ 7180^{ 8300}_{ 6460}$  &  $13.27^{12.64}_{13.74}$  &   80.1  &  0.164  &  $ 7250^{ 8130}_{ 6570}$  &  $13.27^{12.77}_{13.70}$  &   83.8  &  0.147  \\
KX41  &  $ 4650^{ 4820}_{ 4500}$  &  $15.18^{15.03}_{15.33}$  &   72.0  &  1.062  &  $ 5080^{ 5190}_{ 4990}$  &  $15.02^{14.92}_{15.10}$  &   77.4  &  0.874  \\
KX42  &  $ 4680^{ 5050}_{ 4320}$  &  $15.16^{14.82}_{15.51}$  &   54.2  &  0.926  &  $ 5180^{ 5420}_{ 4980}$  &  $14.73^{14.54}_{14.90}$  &   70.6  &  0.459  \\
KX43  &  $15360^{20230}_{ 9990}$  &  $ 9.93^{ 8.72}_{11.82}$  &  143.3  &  0.045  &  $11680^{15030}_{ 9760}$  &  $11.19^{10.08}_{11.97}$  &  109.0  &  0.080  \\
KX44  &  $ 5540^{ 6070}_{ 5110}$  &  $14.41^{14.01}_{14.77}$  &   78.2  &  0.328  &  $ 5550^{ 5980}_{ 5210}$  &  $14.63^{14.45}_{14.90}$  &   68.0  &  0.471  \\
KX45  &  $ 7040^{ 8060}_{ 6370}$  &  $13.36^{12.77}_{13.80}$  &   59.3  &  0.347  &  $ 7140^{ 7990}_{ 6490}$  &  $13.33^{12.84}_{13.75}$  &   62.6  &  0.303  \\
KX46  &  $ 4760^{ 5130}_{ 4390}$  &  $15.08^{14.75}_{15.44}$  &  101.9  &  0.863  &  $ 5260^{ 5520}_{ 5050}$  &  $14.67^{14.45}_{14.85}$  &  133.2  &  0.425  \\
KX47  &  $ 7400^{ 8300}_{ 6670}$  &  $13.14^{12.64}_{13.60}$  &   97.2  &  0.104  &  $ 7720^{ 8730}_{ 6950}$  &  $12.99^{12.46}_{13.45}$  &  107.5  &  0.083  \\
KX48  &  $ 4290^{ 4670}_{ 3850}$  &  $15.53^{15.16}_{16.01}$  &   36.4  &  2.093  &  $ 4970^{ 5180}_{ 4780}$  &  $14.91^{14.73}_{15.09}$  &   51.4  &  0.813  \\
KX49  &  $ 5230^{ 5660}_{ 4860}$  &  $14.67^{14.32}_{14.99}$  &   58.5  &  0.577  &  $ 5570^{ 5900}_{ 5310}$  &  $14.42^{14.17}_{14.63}$  &   69.8  &  0.364  \\
KX50  &  $ 4860^{ 5230}_{ 4500}$  &  $14.99^{14.67}_{15.33}$  &   45.1  &  0.711  &  $ 5350^{ 5630}_{ 5120}$  &  $14.59^{14.37}_{14.78}$  &   58.5  &  0.359  \\
KX51  &  $ 4360^{ 4730}_{ 3950}$  &  $15.47^{15.11}_{15.89}$  &   48.1  &  1.608  &  $ 5020^{ 5240}_{ 4830}$  &  $14.87^{14.68}_{15.04}$  &   67.7  &  0.638  \\
KX52  &  $ 6540^{ 7300}_{ 6000}$  &  $13.68^{13.20}_{14.06}$  &   77.0  &  0.180  &  $ 6730^{ 7440}_{ 6220}$  &  $13.59^{13.15}_{13.94}$  &   83.8  &  0.147  \\
KX53  &  $ 6070^{ 6510}_{ 5700}$  &  $14.01^{13.70}_{14.29}$  &   92.8  &  0.199  &  $ 6000^{ 6400}_{ 5710}$  &  $14.44^{14.14}_{14.56}$  &   73.4  &  0.359  \\
KX54  &  $ 7350^{ 8330}_{ 6610}$  &  $13.17^{12.62}_{13.64}$  &   72.3  &  0.211  &  $ 7570^{ 8540}_{ 6840}$  &  $13.08^{12.55}_{13.52}$  &   78.4  &  0.173  \\
KX55  &  $ 8160^{10020}_{ 6980}$  &  $12.71^{11.81}_{13.40}$  &  175.5  &  0.060  &  $ 7810^{ 8910}_{ 7010}$  &  $12.94^{12.37}_{13.41}$  &  165.8  &  0.068  \\
KX56  &  $ 4900^{ 5300}_{ 4520}$  &  $14.95^{14.61}_{15.31}$  &   96.1  &  0.661  &  $ 5390^{ 5690}_{ 5150}$  &  $14.56^{14.33}_{14.76}$  &  124.7  &  0.335  \\
KX57  &  $ 3750^{ 3980}_{ 3750}$  &  $16.12^{15.86}_{16.12}$  &   41.2  &  5.010  &  $ 4480^{ 4680}_{ 4310}$  &  $15.37^{15.18}_{15.53}$  &   52.5  &  2.538  \\
KX58  &  $ 3750^{ 3930}_{ 3750}$  &  $16.12^{15.92}_{16.12}$  &   28.4  &  4.411  &  $ 4570^{ 4750}_{ 4420}$  &  $15.28^{15.11}_{15.43}$  &   38.9  &  1.824  \\
\hline\hline
\end{tabular}
 \caption{Derived stellar parameters for CWD sample assuming either a H or He
atmosphere.}
 \label{analtab}
\end{table*}

Distance moduli obtained from these fits allow calculation of tangential velocities via the 
SuperCOSMOS proper motion measures.  The distribution of derived tangential velocities is
consistent with expectations for a sample of Disc stars (Evans 1992), and shows no evidence
 of contamination from high velocity halo objects.  A summary of the results of the fitting
procedure is shown in Table~\ref{analtab}, including $T_{\rm eff}$ and $M_{\rm bol}$ with
associated errors and $V_{\rm tan}$ values. Space density values for each object are also
presented in Table~\ref{analtab}; these are discussed in the following section.

An independent check on the validity of our initial CWD sample parameters (SuperCOSMOS 
photometry and astrometry) can be made by comparing known CWD samples with our data
on the RPMD.  Since the RPMD is the original means of population discrimination
it is also interesting to overplot known subdwarf samples on our RPMD to further
address the question of potential contamination.
BRL published observations
of a sample of 110 CWDs, including most of the coolest known degenerates.
Extensive lists of extreme subdwarfs are less easily obtainable.  Ryan (1989) used a RPM
criterion to extract over 1000 subdwarf candidates from the NLTT catalogue.  Accurate
B and R photometry was published for these objects, providing a useful means of 
delineating
the bluer portion of the WD RPMD locus.  Monet et al. (1992) identified a subset of 17 
extreme subdwarfs from their CCD parallax program also involving Luyten catalogue stars.
Although only V, I photometry is published for these stars, we use the photometry published
in Ryan (1989) to define colour transformations allowing the 17 subdwarfs from
Monet et al. (1992) to be plotted
on the $(B-R)$, RPMD planes.  These objects define
a portion of the RPMD marginally red-wards of the faintest CWDs where the most
extreme contaminants may be expected to lie.  
\begin{figure}
\begin{center}
\setlength{\unitlength}{1mm}
\begin{picture}(70,50)
\includegraphics{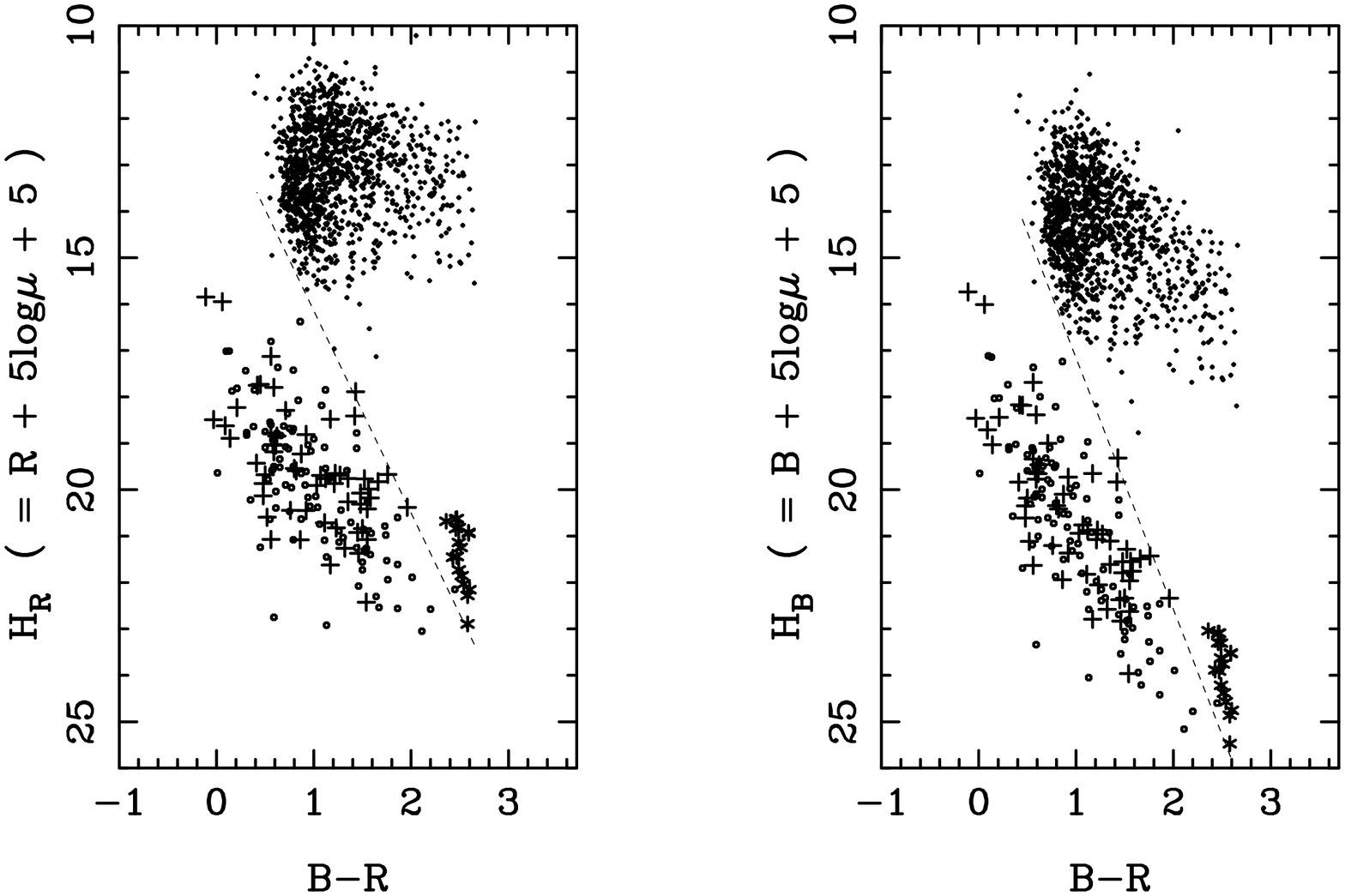} 
\end{picture}
\end{center}
\vspace{2cm}
 \caption{A comparison of various populations on the RPMDs: crosses denote our CWD sample
, circles the BRL CWD sample, asterisks the Monet et al. extreme red subdwarfs
 and dots the brighter Ryan subdwarfs}
 \label{rpmdcomp}
\end{figure}

Figure~\ref{rpmdcomp} shows the two RPMD with the four samples plotted.  There are
several points to be made concerning  this plot.
  Firstly, our CWD sample and the the BRL sample
of previously known CWDs lie on the same region of the diagram, providing further 
confirmation of the validity of our survey procedure.  It can also be seen that the
BRL sample contains redder, cooler stars than our sample. This may be expected since
the BRL sample is rather eclectic and contains some of the coolest WDs known, whereas
our sample is drawn from a rigidly defined survey in a particular ESO/SERC field.  We note
also that the cool portion of the BRL sample does not extend into the portion of the
RPMD beyond our population discrimination cut-off shown in Figure~\ref{samcut}, which
may be interpreted as
indicating that we are not failing to sample portions of the RPMD containing CWDs 
(but see Section~\ref{discuss} below).
Both subdwarf samples lie in clearly distinct regions of the RPMD to our sample, although
the cooler subdwarfs are all too red to directly assess contamination of the CWD
sample from the direction of small RPM. 
  However the subdwarf RPM locus is not predicted to deviate significantly
from a straight line in the CWD colour regime (Evans 1992), and if we take the high
H extent of the two subdwarf samples plotted to be indicative of the limit of the 
extreme subdwarf locus on the RPMD, the dashed lines plotted on Figure~\ref{rpmdcomp}
should be a good guide to the limit of the subdwarf locus for the intermediate colour
range.  It may then immediately be seen that the vast majority of our CWD sample is
safely within the WD region of the RPMD.  The two redder borderline stars have reasonable
spectroscopic confirmation of their WD status (Figures~\ref{KX57specplot} and~\ref{KX58specplot}),
 leaving only one potentially dubious object.

\section{The WDLF}
\label{wdlf}

In order to construct a WDLF, space densities must be calculated for
a survey limited by both apparent magnitude and proper motion.
The CWD survey sample presented in Table~\ref{analtab} 
consists of stars with widely varying intrinsic brightness and tangential velocity,
and is therefore not volume-limited (since for example intrinsically bright
objects are sampled out to greater distances than the coolest, faintest stars).
The standard solution to this problem, Schmidt's (1968, 1975)
$1/V_{max}$ estimator, has been extensively studied with specific
reference to the WDLF (Wood
\& Oswalt 1998).
  The $1/V_{max}$ method assigns each sample object a survey volume defined by 
the maximum distance $d_{max}$ the object could have and satisfy the survey limit criteria.  
For this survey, an object at distance $d$ with proper motion $\mu$ and magnitudes $B$ and
$R$ has $d_{max}$
\begin{equation}
d_{max} =  {\rm min}
 \left[ d\frac{\mu}{\mu_{lim}},d10^{0.2(R_{lim}-R)},d10^{0.2(B_{lim}-B)} \right].
\end{equation}
For the simple case of uniform stellar space density, the survey field solid angle
 $\Omega$ can then be used to calculate the corresponding
$V_{max}$ ($= \Omega d_{max}^{3} / 3$)
.  However, many
 of the objects in our CWD sample yield a $d_{max}$
comparable to the scale height of the Disc, and we follow the
method of Tinney, Reid \& Mould (1993) in generalising our calculation of $V_{max}$
to allow for the truncation of the survey volume by the scale height effect.
In this prescription, the volume is found by integrating over an exponentially decreasing
density law (Stobie, Ishida \& Peacock 1989) with scale height h at galactic latitude b,
yielding the modified expression for the volume out to a distance d:
\begin{equation}
\label{Vcalc}
V = \Omega \frac{h^{3}}{\sin^{3} b} \{2-(\xi^{2}+2\xi+2)e^{-\xi}\},
\end{equation}
where $\xi  = d \sin b/h$ (Equation 9 in Tinney et al. 1993).
Each object is thought of as a sampling of the 
survey volume $V_{max}$, and thus contributes a space density of $V_{max}^{-1}$ to its
particular LF bin.  We adopt the convention of LDM in assigning the uncertainty
in each space density contribution as being equal to that contribution
(ie. $\rm 1\pm1$ objects per sampling).  We may therefore construct a LF simply by summing
the individual space density contributions in each luminosity bin, and the error
obtained by summing the error contributions in quadrature.  A true reflection of the LFs
observational uncertainties should also allow for the uncertainties inherent 
in photometric fits to model atmospheres described in the previous section.
The errors in $M_{\rm bol}$ detailed in Table~\ref{analtab} suggest the introduction
of horizontal error bars in any observational LF is necessary.  When converting
our sample $M_{\rm bol}$ values into luminosity units via
\begin{equation}
M_{bol} = -2.5\log{L/L_{\odot}} + 4.75
\end{equation}
we also calculate the upper and lower 1 sigma luminosity uncertainties for each object.
For a bin containing N objects we then combine eg. the upper luminosity 1 sigma 
uncertainties, $\sigma_{u}$, using
\begin{equation}
\sigma_{U} = \sqrt{\frac{\sum^{N}_{i} \sigma^{2}_{u_{i}}}{N}}
\end{equation}	
to yield an estimate for the horizontal error bar $\sigma_{U}$,
 with an analogous procedure for
the lower luminosity error bounds.  In addition, to give the most realistic 
estimate for the LF in magnitude bins containing a very few objects, we plot
the binned data at the mean luminosity of the objects in the bin, rather than at
the mid-point of that bin.

\begin{figure}
\begin{center}
\setlength{\unitlength}{1mm}
\begin{picture}(70,40)
\includegraphics{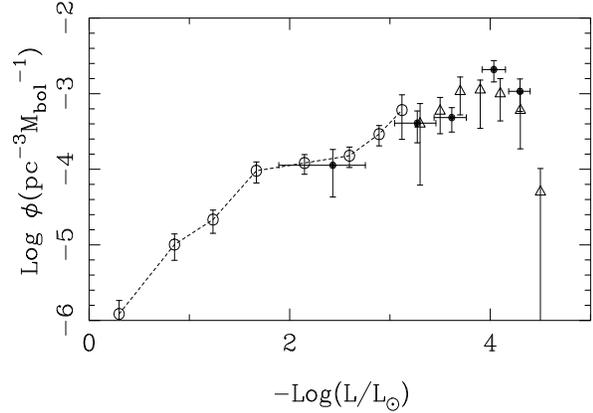} 
\end{picture}
\end{center}
\vspace{2cm}
 \caption{ Observational LF with comparisons: this work (filled circles), hot WDLF 
based on Fleming, Liebert and Green (1986) from LDM (open circles)
and the LRB redetermination of the LDM CWDLF (open triangles)}
 \label{obslf}
\end{figure}

Table~\ref{lf} gives the LF calculated  in this fashion for integer magnitude bins.
Only the cool end of the LF is given here ($M_{\rm bol}>12.25 $), and a field size of 0.0086
steradians and Disc scale height of 300 parsecs were used in calculating space
densities.  Columns 3 and 4 give the plotted (Figure~\ref{obslf})
LF with upper and lower error bounds in
parenthesis.
\begin{table*}
\begin{tabular}{ccccc} \hline\hline
\multicolumn{2}{c} {Bin Center} & & Space Density &  \\
$M_{\rm bol}$ & $-\log{L/L_{\odot}}$ & $\overline{-\log{L/L_{\odot}}} $ & $ \log
{ \{ \left[ \sum (1/V_{max}) \right] \rm M_{bol}^{-1} \} } $ & Number \\ \hline
12.75 & 3.20 & 3.27(3.05,3.45) & -3.39(-3.23,-3.65) & 14 \\
13.75 & 3.60 & 3.62(3.44,3.76) & -3.32(-3.18,-3.51) & 13 \\
14.75 & 4.00 & 4.04(3.92,4.15) & -2.68(-2.56,-2.84) & 20 \\
15.75 & 4.40 & 4.30(4.18,4.40) & -2.97(-2.80,-3.24) &  5 \\
\hline
\end{tabular}
 \caption{Luminosity Function}
 \label{lf}
\end{table*}
The hot WD data point not detailed in Table~\ref{lf} represents the 6 stars with
$(3>-\log{L/L_{\odot}}>2)$.  These hotter objects tend to have large errors
 in fitted $M_{\rm bol}$, and the resulting large horizontal error bars makes 
broader binning appropriate.  It is necessary to choose either a pure hydrogen or 
helium atmosphere for each object to construct the LF.  We use our `best guess'
atmospheres for the LF described here: for each object we choose
the atmosphere with the lower $\chi^{2}$ model fit to the photometry, with the 
exception of objects with $6000>T_{\rm eff_{H}}>5000$ which are deemed to be
occupying the `non-DA gap' at this temperature (BRL) and are therefore
automatically designated a pure H atmosphere.  Note however that the photometry
does not adequately constrain atmospheric composition and this `best guess'  
LF is only one arbitrary realisation of the data (see Section~\ref{discuss} below).

The standard means of assessing the completeness of a sample analysed using the 
$1/V_{max}$ method is to calculate the mean value of the ratio of $V_{obs}$, the
volume out to an object, to $V_{max}$.  A complete survey evenly sampling the survey 
volume should yield $\langle V_{obs}/V_{max} \rangle = \frac{1}{2}$.  This is generally
not the case for published CWD samples, although some authors have incorporated 
completeness corrections into their analyses to account for the effects of the 
original survey incompleteness (OSWH).  The $\langle V_{obs}/V_{max} \rangle$
calculated for this sample is 0.495 or 0.496 choosing either all H or He atmospheres.
From a complete sample containing 58 objects we expect $\langle V_{obs}/V_{max} \rangle
 = 0.5 \pm 0.038$, indicating our sample is consistent with being drawn from
a complete survey.  It should be emphasised however that this result
cannot be regarded as proof of completeness, since clearly an
incomplete survey sample may also exhibit $\langle V_{obs}/V_{max} \rangle \sim \frac{1}{2}$.

The total space density determined from our `best guess' sample is $4.16 \times 10^{-3}$ WDs
per cubic parsec, approximately 25\% greater than that found by LRB.  These results
are certainly consistent, since the simulations of Wood and Oswalt (1998) predict errors
of $\rm \sim50\%$ in total space density estimates from samples of 50 CWDs using the 
$1/V_{max}$ technique and additional uncertainties are introduced by our lack of knowledge
of the WD atmospheric constituents.  Our findings
reiterate that WDs represent only a small fraction ($\sim1\%$)
of the local dynamically estimated mass density.  Interestingly, we do not
 confirm the much higher total WD space densities found recently by two independent
studies.  We note however that these studies (Ruiz \& Takamiya 1995, Festin
1998) make only a tentative claim to detection of a high WD space density due to
the small samples ($\rm N<10$) involved.  A third study (OSWH) searched exclusively for
WDs in CPMBs, and found a total space density of $5.3 \times 10^{-3}$ for these objects.
At this space density, and
given our survey area and the resolution of the COSMOS data, we
would not expect to find any CPMB WDs.  It is therefore not surprising that no object in
 our sample is a CPMB member, since the survey technique is only
sensitive to lone WDs or double degenerate binary systems.

\section{discussion}	
\label{discuss}

An estimate of the Disc age may be obtained by comparing our WDLF with expectations
from theoretical models.  We compare our data with two sets of models, which are available
in the form of curves at integer 1~Gyr Disc age intervals. 
 Since we are fitting to a cut-off in 
space density, the {\em lack} of detected objects beyond our faintest observational
bin assumes added significance.  We can calculate the probability of detecting
zero objects in the next faintest luminosity bin because we have well defined survey
limits: at given faint luminosity the proper motion survey limit is irrelevant
and the survey is sampling a known volume defined by the photometric survey
limits.  This volume was calculated using B and R magnitudes
for very cool WDs from the recent models of Hansen (1998), which combined with the 
photometric survey limits yield a $d_{\rm max}$  for both an H
and an He atmosphere WD.  The minimum $d_{\rm max}$ defines the survey volume at that 
magnitude.  We have assumed the LF at any given luminosity consists of equal
numbers of H and He WDs, in which case the H WD survey volume provides the best constraint
on the LF and is adopted for fitting.  Using Poisson statistics, the probability
of detecting zero objects in a volume V in which a model predicts space density
$\rho$ is simply $e^{-\rho V}$.  We assume the errors on the data points are normally
distributed, and derive best fits by maximising the probability of the
dataset for each curve and comparing the fits for the various Disc ages.

The various inputs to the first set of WDLFs we use, those of Wood, were described
in detail in Wood (1992).  Careful consideration was given to the various inputs to
the WDLF, such as initial mass function, star formation rates and initial-final
mass relation.  The Wood WD evolutionary models for the WDLFs 
have since been updated (Wood 1995, OSWH), and consist of mixed carbon-oxygen core
WDs with hydrogen and helium surface layer masses of $10^{-4}$ and $\rm 10^{-2} M_{*}$
respectively, and also utilise revised opacities and neutrino rates.
It is these more recent model WDLFs that are used here.

\begin{figure}
\begin{center}
\setlength{\unitlength}{1mm}
\begin{picture}(70,40)
\includegraphics{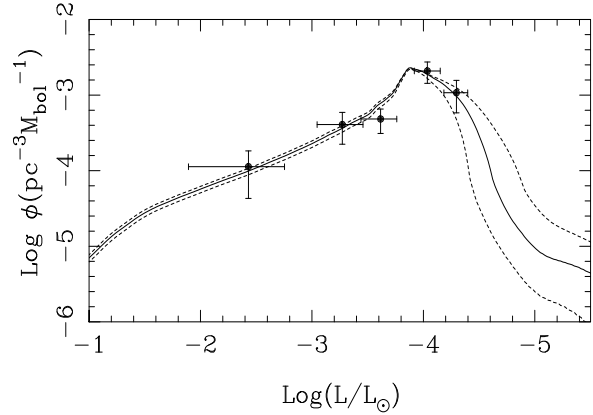} 
\end{picture}
\end{center}
\vspace{2cm}
 \caption{The Wood WDLF with a Disc age of 9~Gyr -- the best fit to our `best
 guess' LF. 8~Gyr (below) and 10~Gyr (above) curves are also shown in dashed lines.}
 \label{woodfit}
\end{figure}

\begin{figure}
\begin{center}
\setlength{\unitlength}{1mm}
\begin{picture}(70,40)
\includegraphics{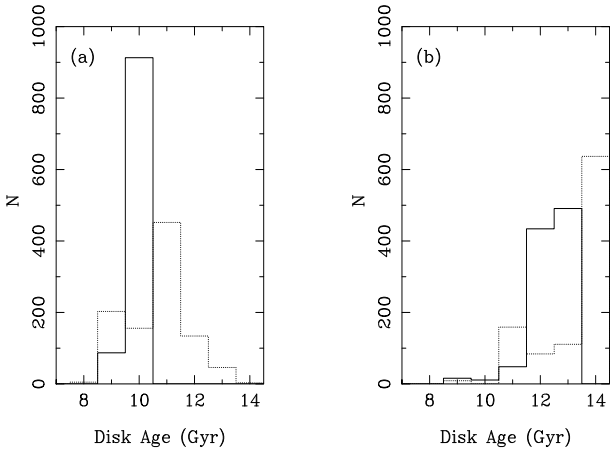} 
\end{picture}
\end{center}
\vspace{2cm}
 \caption{Results of constructing WDLFs from our sample by randomly assigning H or He atmospheres.
Full line histograms denote 1 $M_{\rm bol}$ bins, dotted lines  0.75 $M_{\rm bol}$ bins.
For each bin size 1000 realisations of the WDLF were constructed and fit to the models - the number
of times each Disc age was the best fit is displayed. Graph (a) shows fits to the Wood models, graph
(b) fits to GB.}
 \label{agehists}
\end{figure}

The best fit of our `best guess' LF  to the Wood model WDLFs is for a Disc age of 9~Gyr,
and is shown in Figure~\ref{woodfit}. 
Although this gives a reasonable first indication of the Disc age implied by our sample,
further investigation is necessary since the photographic photometry fitting procedure
used to estimate $T_{\rm eff}$ (as described in Section~\ref{samanal})
does not reliably constrain the atmospheric constituent of the stars in our sample.
We have addressed this question by constructing a large number of WDLF `realisations'
 from our sample
data, each time giving each star a 50\% probability of having either a H or He atmosphere.
Every resulting WDLF was fit to the models and the best fit Disc age recorded.  We have also
used this analysis to assess the effect of binning on the fitting procedure.
Figure~\ref{agehists}~(a) displays the results of fits to 1000 realisations binned in 
1.0 $M_{\rm bol}$ bins and a further 1000 in 0.75 $M_{\rm bol}$ bins.  These results
give a fairer indication of the Disc age and associated errors than Figure~\ref{woodfit}, 
which is effectively one arbitrarily picked realisation.  

\begin{figure}
\begin{center}
\setlength{\unitlength}{1mm}
\begin{picture}(70,40)
\includegraphics{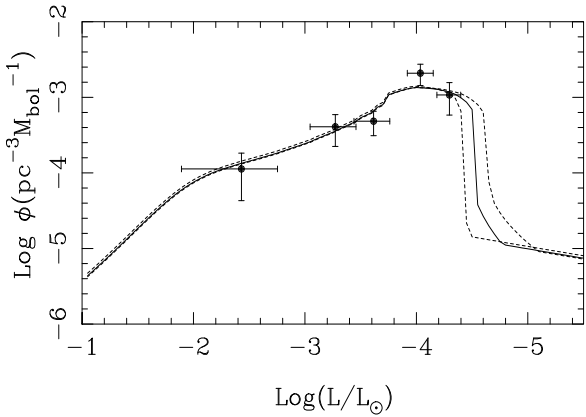} 
\end{picture}
\end{center}
\vspace{2cm}
 \caption{The best fit of our `best guess' LF to the GB WDLF: a Disc age
of 11~Gyr with `adjacent' Disc age LFs shown in dashed lines as in Figure~\ref{woodfit}.} 
 \label{garfit}
\end{figure}

The second set of theoretical WDLFs we use are described in Garc\'{\i}a-Berro
 et al. (1997) (henceforth GB models).
These LFs include the expectation that the progenitors of the faintest WDs
 are likely to have been massive stars since these stars evolve more quickly and
the resulting (Oxygen-Neon) massive WDs also cool faster.  These models also include a
predicted delay in Carbon-Oxygen WD cooling induced by the separation of C and O
at crystallization (Hernanz et al. 1994).
The incorporation of these considerations into the theoretical WDLF
 leads to a broader predicted peak at the tail of the
LF.  We find a best fit of 11~Gyr to our sample, as shown in Figure~\ref{garfit}.
Again, we have investigated the effect of our poor knowledge of our samples atmospheres
in the same way as above.  The results, shown in Figure~\ref{agehists}~(b), highlight the
effect that variations in binning can have on the fits for this second
set of models.  

  The major contributor to this problem seems to be that the model curves
are sufficiently indistinct in the region of our sample 
dataset (as can be seen clearly in Figure~\ref{garfit})
that changes in observational LF binning can have a significant bearing on the 
result of the fit.  This effect is apparent in Figure~\ref{agehists}~(b), where
a bin size of 0.75 $M_{\rm bol}$ yields a strong preference of a Disc age of 14~Gyr, in 
contrast to the 12-13~Gyr Disc age selected by the other binning regime (note also that
14~Gyr was the oldest curve available for fitting).
Although it is difficult to resolve this matter satisfactorily with the current data set, 
there are two pertinent points to raise.
The problem does not apply to the Wood models, the fits to which
are extremely difficult to dislodge from the 9--11~Gyr region indicated by 
Figure~\ref{agehists}~(a). Secondly, Wood and Oswalt (1998), as a result of their Monte Carlo 
analysis of the WDLF, recommend choosing a binning in which the crucial
lowest luminosity bin contains $\sim5$ objects; giving a 
reasonable compromise between the requirements of
good signal to noise in the final bin and having that bin as faint as possible to provide
maximum information on the position of the cut off.  This means a bin size of 
 1 $M_{\rm bol}$ for our data set, for which the fitted Disc age is well constrained for
both models.

A brief investigation of the effect of altering the Disc scale height in Equation~\ref{Vcalc}
revealed that for any scale height between 250 and 400 pc the alterations in space densities
for a LF binned as in Table~\ref{lf} are restricted to a few
hundredths in $\log \phi$.
Variations at this level have a negligible effect on the fits to model
LFs.
This behaviour is to be expected , since it is objects with large $d_{max}$
which are most affected by changes in the scale height,
 and these make the smallest contributions to the space density when using the
$1/V_{max}$ prescription.

An overview of all the various fittings to both sets of models points firmly to a Disc age
between 9 and 11~Gyr using the Wood models, and an age of 12-14~Gyr using the GB models.
This discrepancy is expected (LRB, Garc\'{\i}a-Berro et al. 1997) and indicates the extent
of errors introduced by uncertainties in the WD crystallisation process; the Wood model
Disc age should be a reliable lower limit however, and Figure~\ref{agehists} demonstrates 
the difficulty in obtaining a Disc age of less than 9~Gyr from our data.
  We note that the Wood models seem
to represent our data better, the GB models not following the peak in the observed
LF.  Wood (1992)
reported that the uncertainties in the inputs to model WDLFs lead to a further $\sim1$~Gyr
contribution to the total error. 
Further
insight into the error associated with the Disc age may be gained by
considering the extremities of the distribution of atmosphere types
within the sample. The 50\% probability assigned to the H and He
atmosphere is adopted in the absence of strong
evidence for a dominant atmospheric type amongst the crucial coolest
WDs (see eg. the coolest bin in Figure 1 of BRL), and has no physical
basis.  Adopting all H atmospheres leads to high space density
estimates in the final LF bins and arbitrarily old Disc age estimates.
More interestingly, a LF composed of all He atmosphere WDs is still
incompatable with Disc ages below 8 Gyr, regardless of the model WDLF
used.  This reiterates the important point that the cosmologically
interesting lower Disc age limit appears to be  8~Gyr, and that even a Disc
as young as this must be considered unlikely (Figure~\ref{agehists}).
Our adopted Disc age estimation is therefore 
$10^{-1}_{+3}$~Gyr.

There are at least two
possible effects arising from our present lack of extensive follow up observational data
 that could affect the derived Disc age. First, as may be seen in Figure~\ref{spechplot}
there is a small group of objects
lying just beyond our RPMD cut-off that have not been included in our CWD sample.  
Although the distribution of known WDs on the RPMD (the BRL objects in 
Figure~\ref{rpmdcomp}) indicates a population of CWDs is not expected in this region, the
reason for this may be that many of the known CWDs have been themselves selected on the 
basis of RPM criteria (Luyten 1970). It may be that the region of the RPMD just beyond our
cut-off has not been adequately investigated before, since in the presence of noisier data
from blink comparators and eye-measure photometry it would be hopelessly confused.
These considerations argue strongly for a wholesale spectroscopic survey using multi-fibre
instruments of the entire population below $H_{R}=19$ for complete confidence that our
sample does not exclude any CWDs.  For the present, the possibility that a few more CWDs
exist in our catalogue but do not satisfy our RPMD survey criteria cannot be ruled out.
  Such objects would certainly be very 
cool, resulting in a higher Disc age estimate.
The second effect concerns the question of mass.  The difficulty is amply demonstrated
by the case of CWD ESO 439-26 (Ruiz et al. 1995), which was observed to have a luminosity
fainter by 1 magnitude than the WDLF cut off.  Analysis of the object's optical energy
distribution in conjunction with a measure of trigonometric parallax allowed the authors
to conclude that it's low luminosity was in fact due to its large mass, or small radius.
Again, ideally it would be desirable to obtain parallaxes and
CCD optical and IR photometry of our entire sample, obviating the need to assume a mass of
$0.6M_{\odot}$ (and allowing the surface composition to be constrained).  

In summary, our WD sample has passed every test for completeness  applied to it. 
The calculated space density of WDs is slightly higher than that for the LDM sample,
 however we do not detect the much higher total space densities found by more recent authors
(Ruiz 1995, Festin 1998).  Our WDLF yields a Disc age estimate of $10^{-1}_{+3}$~Gyr, 
older but still consistent with  previous estimates (Winget 1987, LRB, OSWH). In
the context of current cosmochronometry, our Disc age estimate is consistent with current 
Globular cluster age estimates of 13-14Gyr (Vandenberg 1998).  Finally, when
combined only with a conservative 1~Gyr value for the halo--Disc formation interval 
(Burkert, Truran \& Hensler 1992, Pitts \& Tayler 1992) and a
further 1 Gyr for the big bang--halo formation interval, a 10~Gyr Disc excludes $\Omega = 1$,
$\Lambda = 0$ cosmologies based on current estimates for $H_{\circ}$ of 60-80 (Freedman 1998).

\section{Conclusions}

\begin{enumerate}
\item Using a large collection of COSMOS/SuperCOSMOS digitised Schmidt plate data in 
ESO/SERC field 287 we have extracted a sample of proper motion objects.  Number counts
indicate this sample is complete down to the survey proper motion limit, which was 
chosen with care to exclude contaminant spurious proper motion measures.

\item A sample of cool white dwarfs have been culled from our proper motion objects using
the reduced proper motion technique.  By overplotting samples of known extreme subdwarfs
we show our CWD sample is unlikely to be contaminated
by other stellar population groups. We have confirmed the WD status of a number of our
sample with AAT spectroscopy.  The sample passes the  $(V/V_{max})$ completeness test.

\item We calculate a total WD space density of $4.16 \times 10^{-3}$ WDs
per cubic parsec using Schmidt's $(1/V_{max})$ method.  Careful comparison
of luminosity functions constructed from our sample and theoretical models indicate
an age for the local Galactic Disc of $10^{-1}_{+3}$~Gyr, older than previous estimates using
this technique.
\end{enumerate}	

\section*{Acknowledgments}

We would like to thank M. Wood and E. Garc\'{\i}a-Berro for access to theoretical
WDLFs, and P. Bergeron, P. Hauschildt and B. Hansen for supplying model atmosphere predictions.
Thanks also to S. Ryan for supplying lists of subdwarfs, and to Andy Taylor for useful discussions
concerning statistics.
This work would not have been possible without the time and expertise of the SuperCOSMOS and
UK Schmidt Library staff.  Particular thanks to Harvey MacGillivray, Mike Read and Sue Tritton.
 Richard Knox acknowledges a PPARC postgraduate studentship.

\appendix

\section{Reduced Proper Motion Diagrams}
\label{rpms}

The reduced proper motion (RPM) is defined by
\begin{equation}
\label{rpm}
 H = m + 5\log_{10} \mu + 5
\end{equation}
where m is apparent magnitude and $\mu$ proper motion.
A reduced proper motion diagram (RPMD) is a plot of colour against RPM.  It is
a powerful way of combining proper motions and photometry to distinguish 
stellar population groups.  Equation~\ref{rpm} can be re-written using the 
relationships $ m = M - 5 + 5\log d$ and $\mu = V_{T} / 4.74d$ to give
\begin{equation}
 H = M + 5\log_{10} V_{T} - 3.379
\end{equation}
where  M is the absolute magnitude, $V_{T}$ the transverse velocity (in $\rm km s^{
-1}$) and d the distance. Since M and $V_{T}$ are both intrinsic properties of
the star, so too is H.

 To see the significance of H, suppose that every star
had an identical $V_{T}$; H would then clearly be simply M plus a constant and
 the distribution of a particular population group in H at a
 particular colour would depend solely on the spread of 
the populations' colour-magnitude relation at that 
colour.  Of course there is a distribution in $ 5\log_{10} V_{T}$ for each 
population, but since the tangential velocities are distributed 
around a most probable value the RPM serves as an estimate of M, ie. 
 $M = a + bH$ (a and b constants).  The resulting locus for each population 
 in the RPMD is
 then the convolution of its colour magnitude distribution with its 
$ 5\log_{10} V_{T}$ distribution over the diagrams colour range.
 To allow population discrimination in some colour region we
 therefore
 only require that the various population loci  do not overlap significantly
 in that region of the RPMD.  In effect the RPMD  is analogous
 to the Hertsprung-Russell diagram, and
in both plots the white dwarf population is quite distinct in most colours.

\vfill
\bsp

\end{document}